\documentclass[aps,prx,floats,floatfix,twocolumn,longbibliography]{revtex4-1}

\usepackage{graphicx}
\usepackage{epstopdf}
\graphicspath{{./}{./Figs/}{./figs/}}

\usepackage[percent]{overpic} 

\usepackage{latexsym}
\usepackage{amsmath}
\usepackage{amssymb}
\usepackage{bm}
\usepackage{wasysym}

\usepackage{color}
\usepackage[colorlinks,linkcolor=blue,urlcolor=blue,bookmarks=true,pdftitle={}]{hyperref}
\usepackage{flafter}

\newcommand{\im}{\mbox{Im}}
\newcommand{\re}{\mbox{Re}}

\newcommand{\bra}[1]{\left\langle #1 \right|}
\newcommand{\ket}[1]{\left| #1 \right\rangle}

\newcommand{\Braket}[2]{\left\langle #1 \middle| #2 \right\rangle}
\newcommand{\BraKet}[3]{\left\langle #1 \middle| #2 \middle| #3 \right\rangle}
\newcommand{\Ketbra}[2]{\left| #1 \middle\rangle \middle\langle #2 \right|}
\newcommand{\KetBra}[3]{\left| #1 \middle\rangle #2 \middle\langle #3 \right|}

\newcommand{\ola}{\protect\overleftarrow}
\newcommand{\ora}{\protect\overrightarrow}

\newcommand{\be}[1]{\begin{eqnarray}{\label{e#1}}} 
\newcommand{\beq}{\begin{eqnarray}}
\newcommand{\eeq}{\end{eqnarray}} 

\newcommand{\hide}[1]{}
\newcommand{\Eq}[1]{\textcolor{blue}{{Eq.}\!\!~(\ref{#1})}}
\newcommand{\App}[1]{\textcolor{blue}{{Appendix}\!~\ref{#1}}}
\newcommand{\Sec}[1]{\textcolor{blue}{{Section}\!~\ref{#1}}}
\newcommand{\Fig}[1]{\textcolor{blue}{Fig.}\!\!~\ref{#1}}
\definecolor{myred}{rgb}  {0.5,0.0,0.0}

\newcommand{\hrefl}[2]{\href{#2}{(#1)}}
\newcommand{\sect}[1]{{\bf #1.--}}

\newcommand{\+}{^\dag}

\begin{document}

\title{Localization due to topological stochastic disorder in active networks}

\author{Dekel Shapira, Dganit Meidan, Doron Cohen}

\affiliation{Department of Physics, Ben-Gurion University of the Negev, Beer-Sheva 84105, Israel}

\begin{abstract}
An active network is a prototype model in non-equilibrium statistical mechanics.
It can represent, for example, a system with particles that have a self-propulsion mechanism.
Each node of the network specifies a possible location
of a particle, and its orientation.
The orientation (which is formally like a spin degree of freedom) 
determines the self-propulsion direction. 
The bonds represent the possibility to make transitions: 
to hop between locations; or to switch the orientation.
In systems of experimental interest (Janus particles), 
the self-propulsion is induced by illumination.
An emergent aspect is the {\em topological stochastic disorder} (TSD).   
It is implied by the non-uniformity of the illumination.
In technical terms the TSD reflects the local 
non-zero circulations (affinities) of the stochastic transitions.     
This type of disorder, unlike non-homogeneous magnetic field, 
is non-hermitian, and can lead to the emergence of a complex relaxation spectrum. 
It is therefore dramatically distinct from the conservative Anderson-type 
or Sinai-type disorder.  
We discuss the consequences of having TSD. 
In particular we illuminate 3~different routes to under-damped relaxation, 
and show that localization plays a major role in the analysis.   
Implications of the bulk-edge correspondence principle are addressed too.  
%
%
\end{abstract}

\maketitle

\section{Introduction}
\label{sec:intro}

Gas that consists of particles that perform self-propelled stochastic motion
is a novel paradigm in statistical mechanics
~\cite{Romanczuk2012,Thompson2011,Katz2011,Marchetti2013}.
Many publication have focused on the implied
hydrodynamic properties of such active matter.
For example its pressure \cite{SolonFilyBaskaranEtAl2015}, 
odd viscosity (for spinning-particles) \cite{banerjee2017odd}
or the similar ``cross-stream migration'' behavior \cite{Katurieaao1755} 
\hrefl{see movie}{https://www.youtube.com/watch?v=oJ4FMc0J-l4}.
The aspect that we would like to address is not related 
to the non-equilibrium steady state (NESS) of the active system, 
but rather to its relaxation dynamics.

To be specific we highlight, as an example,  
a system that consists of Janus particles 
~\cite{Walther2013,Wheat2010,Menzel2015}.
Those are spherical-like nano-particles,
coated at each of their two hemispheres with different materials.
Immersed in solution, and radiated with light,
they produce self-propelled motion.
If a microgear is placed in such active solution it will rotate~\cite{maggi2015} 
\hrefl{see movie}{https://www.youtube.com/watch?v=99ockGMSpHI}.

\begin{figure}
\centering
\begin{overpic}[width=6cm]{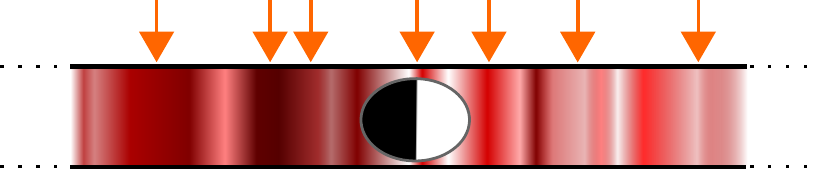} 
\put(-13,5){(a)}
\end{overpic}
\begin{overpic}[width=6cm]{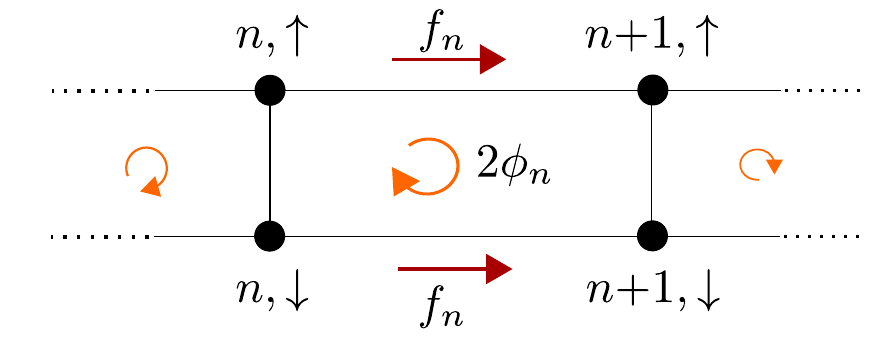} 
\put(-13,18){(b)}
\end{overpic}
\begin{overpic}[width=6cm]{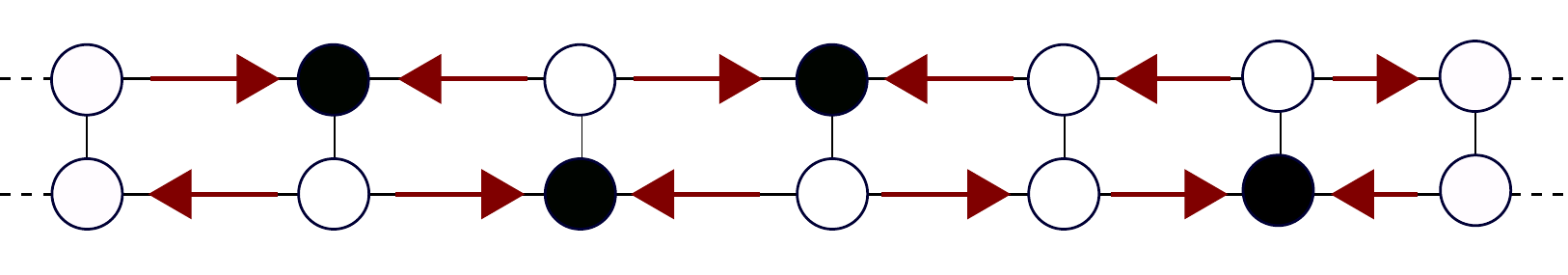} 
\put(-13,7){(c)}
\end{overpic}
\caption{{\bf Model geometry.} 
{\bf (a)} A Janus particle in 1D random environment.
The non-uniform illumination (arrows) induce self-propulsion 
in the direction of the head (pained white). 
The orientation of the head is called below `polarization'. 
{\bf (b)} The lattice modeling of the system. 
Each node represent possible location ($n$) and possible 
polarization ($s=\uparrow,\downarrow$). 
One tile of the lattice is plotted. 
The system is composed of $N$ tiles with periodic boundary conditions.
The transition rates along the vertical bonds are $w{=}1$ 
in both directions, representing random flips of polarization. 
The horizontal bonds are biased:   
the stochastic field there ($\mathcal{E}$) is written 
as the sum of a drift ($f_n$) and a self-propulsion term ($\phi_n$).  
The latter reflects that we are dealing with an active network.
{\bf (c)} An illustration for a representative segment 
of the lattice. The black sites are those 
that serve as sinks for the stochastic flow 
in the presence of strong disorder. 
They support the floor-level relaxation modes. 
}\label{f1}
\end{figure}

\Fig{f1}(a) shows a caricature of a Janus particle in one-dimensional system.
In this caricature the particle can
be anywhere along the horizontal axis~($n=\text{integer}$),  
and can either face to the right or to the left ($s=\uparrow,\downarrow$).
The illumination provides the self-propulsion mechanism. 
For a given orientation the particle executes a biased stochastic random walk: 
the motion is biased to the right (left) if the particle is facing the right (left). 
Formally the orientation of the particle is like a spin degree of freedom, 
and below we refer to it as ``polarization".  

Janus particles that form such active system,
can be placed in a disordered background environment \cite{Morin2017}.
This kind of disorder is derived from a potential, and as 
such is termed {\em conservative}.
Irrespective of that, the illumination might be non-homogenous.
This non-homogeneity provides a new type of disorder.
As explained below such type of inherently  different disorder is {\em topological} 
rather than conservative.

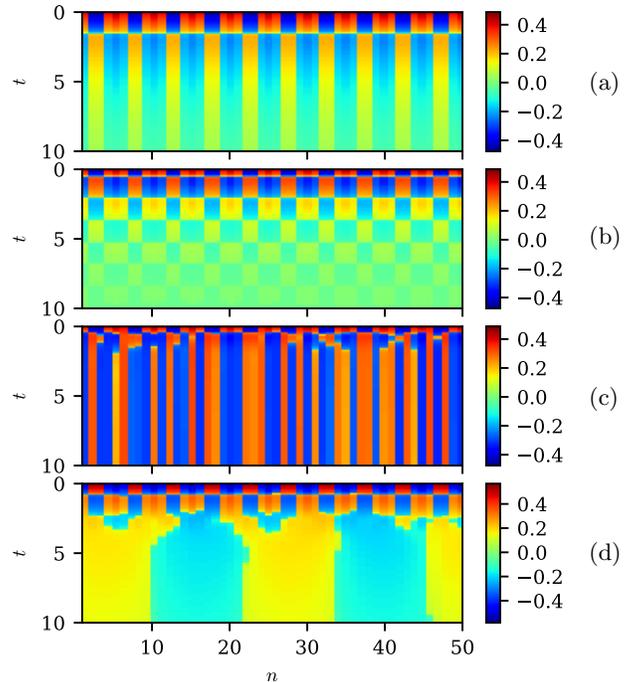
\begin{figure}
\begin{overpic}[width=8cm]{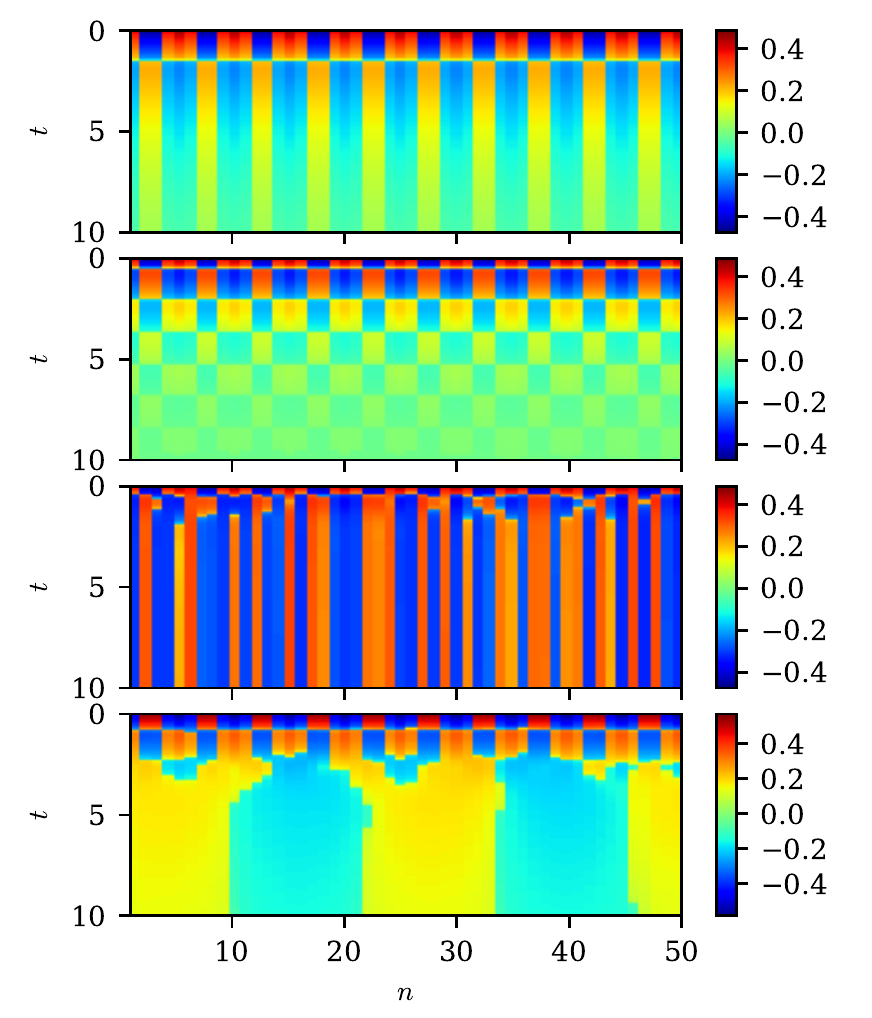} 
\put(85,86){(a)}
\put(85,64){(b)}
\put(85,41){(c)}
\put(85,19){(d)}
\end{overpic}

\caption{ 
{\bf Simulation of polarization as a function of time.} 
The average polarization~$D_n$ at each location~$n$ is color-coded,  
and imaged as a function of time~$t$.
In all the panels the polarization of the initial perpetration 
is modulated with wavenumber~${k=2 \pi/5}$, 
namely, ${p_{n,s} \propto 1 {\pm} \cos(kn)}$. 
Only $50$ sites are displayed.   
{\bf (a)}~Over-damped oscillations for ${\phi=0.5}$  
are observed since the self-propulsion 
is below the critical threshold. 
{\bf (b)}~For ${\phi=2}$, which is above the threshold,  
one observes under-damped relaxation 
due to topological symmetry breaking.
{\bf (c)}~Once disorder is added, the non-uniform NESS pattern 
overtakes almost immediately. Parameters here are as in \Fig{f2}.
{\bf (d)}~In the latter case we provide an image of the time derivative $\dot{D}_n$, 
hence one can resolve how the under-damped relaxation is blurred.  
}\label{f10}
\end{figure}

\sect{Dynamical simulation}
In order to motivate the study of the relaxation dynamics 
let us discuss the simulation that is provided by \Fig{f10}. 
In an actual experiment the particles occupy a two-dimensional strip, 
and some ordering field encourages them to have either left or right polarization. 
We prepare, at time $t{=}0$, a solution that has a uniform density 
of Janus particles, but with modulated polarization. 
In the figure, the color code is such that red and blue regions  
occupy right-polarized (${s=\uparrow}$) and left-polarized (${s=\downarrow}$) particles respectively.
The dynamics is formally described by a stochastic rate equation
\be{1}
\frac{d}{dt}\bm{p} \ \ = \ \ \bm{W} \, \bm{p}  
\eeq
where $\bm{p}$ is vector of probabilities, 
and $\bm{W}$ is a real asymmetric matrix 
that is determined by the transition rates.  
Due to the illumination, the right (left) polarized particles 
execute self-propelled stochastic motion that 
is biased to the right (left) direction.   
In the absence of disorder we see that the relaxation of the system 
becomes under-damped instead of over-damped if the 
strength of the illumination ($\phi$) exceeds a critical value.
We also see that in the presence of disorder both the steady state 
and the relaxation scenario become very different.

\sect{Complex spectrum}
In order to understand the relaxation of a stochastic system, 
as in the simulation discussd above, 
we have to inspect its relaxation modes. These are the 
eigenvectors of the real $\bm{W}$ matrix of \Eq{e1}.
They decay in time as ${\exp(-\lambda_r t)}$, 
where the $\{-\lambda_r\}$ are the associated eigenvalues.
The relaxation becomes under-damped, meaning that it exhibits oscillations in time, 
if some of the $\lambda_r$ acquire an imaginary part.   
Formally we are dealing with the physics of real non-hermitian matrices, 
where the spectrum might become complex 
~\cite{Berry2004,bender2007making,Hatano1996,Hatano1997,Shnerb1998,Feinberg1999,KAFRI2004,KAFRI2005,Amir2016,HurowitzCohen2016,HurowitzCohen2016a,Gong2018}.

\sect{Observations}
Our objective is to provide a precise quantitative framework  
for the analysis of relaxation in active networks.
In particular we are interested in the implications of disorder, 
and the possible manifestation of
localization effect in the relaxation modes of such systems.
We are going to explore different routes 
towards under-damped relaxation. The traditional way for getting 
under-damped relaxation is to introduce bias that drifts 
the particle in one direction. 
In the present study we are going to discuss 
the possibility to get under-damped relaxation
due to self-propulsion. 
We are going to see that uniform illumination requires 
a finite  threshold value, unlike non-uniform illumination
which induces complexity as soon as it is introduced.   
We shall see that stochastic disorder can either 
stabilize or destabilize the over-damped modes,  
depending whether it is conservative or topological.

\sect{Active network}
We deal with particles that execute self-propelled motion. 
The formal modeling is in terms of a network:  
for our specific configuration it is the $N$-tile lattice of \Fig{f1}b.  
The network is described by the matrix $\bm{W}$ of \Eq{e1}.
On each bond $b$ we define a stochastic field~$\mathcal{E}_b$
that indicates how the transitions are biased.
We use the notation $\mathcal{B}_n$ for the circulation 
of $\mathcal{E}_b$ around the $n$-th tile of the network.  
{If all the circulations $\mathcal{B}_n$ are zero,
aka {\em detailed balance}, 
we say that the system is {\em conservative}.} 
In such system the NESS is a canonical equilibrium state 
with zero currents. 
Non-zero circulations are called affinities, 
and can drive NESS currents.

\sect{Topological disorder}
In the present work we consider a new type of disorder that 
we call {\em topological stochastic disorder} (TSD).
It arises due to having a random $\mathcal{B}_n$, 
and therefore constitutes a generic feature of active network.
{This is in contrast with having random $\mathcal{E}_{b}$ with zero $\mathcal{B}_n$, 
which generates conservative disorder}.
{The topological disorder is a novel}
type of disorder which is very different from (say) random 
magnetic field~\cite{Chan1996}, because it breaks the 
hermiticity of $\bm{W}$, leading to the appearance of a complex spectrum.   
{Such type of disorder, that arises due to non-zero circulations, 
requires more than one-channel, or more than one dimension, 
hence the term ``topological''. 
In two dimensions, the continuum-space limit of a stochastic 
random walk process leads to a diffusion equation. 
The presence of TSD implies the appearance of a {\em solenoidal} random drift field.  
The effect of the latter has been discussed by \cite{Kravtsov1985},  
where it has been shown that it leads to a slightly 
super-diffusive spreading.}

\sect{Floor level}
{Strong disorder tend to attracted the system into sinks, 
as illustrated in \Fig{f1}(c). 
In the case of conservative disorder those sinks are stable minima 
of the stochastic potential. In the case of TSD the stability 
of the sinks is marginal. Either way, after a transient time 
the probability distribution get concentrated within this floor level.  
The term {\em floor level} refers to this set of sinks, 
that supports the slow-relaxation modes of the network.}

\subsection{Related studies}
\label{sec:background}

\sect{Topological NESS}
The topological nature of the NESS for the model that we are considering, 
without disorder, has been discussed in~\cite{Murugan2017,Vaikuntanathan2017}, 
and a connection has been established with the Su-Schrieffer-Heeger model 
following the work of~\cite{Kane2014} on topological boundary modes in isostatic lattices. 
In the present work we are not considering the NESS, but rather the relaxation modes, 
and their bulk localization properties due to disorder.  
We note also that the NESS of similar non-disordered quasi-one-dimensional models 
has been investigated in the context of traffic with exclusion rules,  
see for example~\cite{Reichenbach2006,Helbing2001}. 
The main focus in the latter case was the formation 
of a polarization wall due to the entering and the exiting
rates at the boundaries.

\sect{The Sinai model}
The minimal model for stochastic motion in disordered lattice is due to Sinai
~\cite{Sinai1983,Derrida1983,BOUCHAUD1990,BOUCHAUD1990a}, 
aka random walk in random environment. 
In Sinai model the particle can hope between neighboring sites of a Chain, 
and the rates of transition are random numbers.
This leads, in the absence of bias, to sub-diffusion.
Above some critical bias the drift velocity becomes non zero, 
which is known as the {\em sliding transition}.

\sect{The Hatano-Nelson model}
The asymmetry of the~$\bm{W}$ matrix in Sinai's model can be gauged away, 
hence it is similar to an Hermitian matrix, and possesses a real spectrum.  
However, this is not the case if the chain is closed into a ring 
(i.e. imposing periodic boundary conditions). It has been realized
~\cite{Hatano1996,Hatano1997,Shnerb1998} that above some critical bias 
the spectrum becomes complex, and the term {\em delocalization transition} 
has been coined. The subtle relation to the sliding transition in Sinai's model 
has been investigated in a later study~\cite{HurowitzCohen2016}.

\sect{Anderson localization}
The analysis of Sinai's model, which describes random walk 
in a {\em non-active} disordered environment, 
is strongly related to studies of Anderson-type localization. 
It is customary to distinguish between two types of disorder, 
so called Type-I and Type II \cite{Dyson1953,Ziman1982}.   
Roughly speaking the former arises from on-site disorder, 
while the latter arises from bond disorder, in the senses 
of resistor network models \cite{Alexander1981}.

\subsection{Outline}
\label{sec:outline}

We provide a precise definition of the minimal model in \Sec{sec:model}, 
where we also review the different types of disorder that can arise in a network.    
We highlight in \Sec{sec:routes} the different routes to complexity. 
\Sec{sec:symmetry-breaking} provides a detailed account for 
the breakdown of reality due to self-propulsion in the absence of disorder. 
\Sec{sec:TSD} explains how the spectrum is affected by the introduction 
of TSD, and why the threshold for complexity diminishes due to the disorder.   
In \Sec{sec:characterization-eigenmodes} we introduce several measures 
for the characterization of a relaxation mode,  
that help to gain a deeper insight into the spectrum. 

Conservative disorder is responsible to the robustness of reality, 
meaning that eigenvalues remain real even if 
not-too-strong circulations are introduced.   
In contrast TSD leads to complexity via topological mixing.
But there is a twist: we observe in \Sec{sec:floor-level}
that strong stochastic disorder,
irrespective of its nature, induces ``lattice dilution", 
leading to the formation of a {\em floor-level}. 
Consequently the effective dimensionality of the lattice reduces, 
and a robust reality is gained within this floor-level.
The topological-index perspective of the disorder, 
and its connection to the floor level phenomenology 
is further discussed in \Sec{sec:topo}.

\section{The minimal model}
\label{sec:model}

\begin{table*}
\begin{tabular}{|l|l|l|l|}
\hline
{\bf Type} & {\bf aka} & {\bf Relevant models}  &  {\bf Comments} \\   
\hline
Type-I & Diagonal disorder & Anderson model (random potential) &  Might have a mobility edge  \\   
\hline
Type-II & Bond disorder & Debye model; Random-resistor-network & Might lead to a percolation transition \\  
\hline  
MFD & Phase Disorder & Anderson with random vector potential  & The topological aspect is not pronounced  \\  
\hline 
CSD & - & Sinai model of random stochastic transitions & Reduces to type-I via a gauge transformation \\
\hline
RSD & Sign Disorder & Random excitatory and inhibitory connections  & Non-hermiticity cannot be gauged away \\
\hline
TSD & - & Generic active networks  & Non-hermiticity is of topological origin \\
\hline
\end{tabular}
\caption{Different types of disorder.
}\label{t1}
\end{table*}

We consider a minimal configuration for a self-propelled particle in a random environment.
Namely, we assume that the dynamics takes place on a quasi one-dimensional grid, see \Fig{f1}(b).
If the particle is facing the right, we say it has `right' polarization. 
If its black-white orientation is opposite, we say that it has `left' polarization. 
Accordingly, its states $\ket{n,s}$ are defined in terms of position ($n=\text{integer}$) 
and spin coordinate ($s=\uparrow,\downarrow$).
Below we refer to the system as `lattice' that consists of `sites'. 
Each two sites with the same index~$n$ form a `cell', 
and two adjacent cells, along with their connecting bonds, form a `tile'.
The dynamics is described by the rate equation, \Eq{e1}.  
The off-diagonal elements of the $\bm{W}$ matrix are the transition rates~$w$ (with an appropriate bond index).  
The diagonal elements~$-\gamma$ (with an appropriate state index) 
are implied by conservation of probability (the sum of each columns has to be zero).  
The matrix $\bm{W}$ is given explicitly in \App{sec:matrix-elements}.
The rate of transition between two sites, connected
by a bond ($b$), is characterized by a stochastic field 
\beq
\label{eq:epsilon-b}
\mathcal{E}_b \ \ = \ \ \ln\left(\frac{w_{\ora{b}}}{w_{\ola{b}}}\right)
\eeq
whose sign indicates the preferred sense of transition. 
Thus the rates on a given bond can be written as ${w_b\exp(\pm \mathcal{E}_b/2)}$.
In the geometry of \Fig{f1}(b), the vertical bonds represent random flip of orientation, 
and therefore are characterized by a zero stochastic field. 
In contrast the horizontal bonds are biased. 
The stochastic field on the bond ${b=(n,s)}$ that connects 
node $\ket{n,s}$ to $\ket{n{+}1,s}$ is conveniently 
written as a sum of drift and self-propulsion terms, namely, 
\beq \label{eq:epsilon-equals-topo-field}
\begin{aligned}
\mathcal{E}_{n,\uparrow}   \ &= \ f_{n} + \phi_n \\
\mathcal{E}_{n,\downarrow} \ &= \ f_{n} - \phi_n 
\end{aligned}
\eeq
The activity of the network is reflected in having non-zero circulations,  
aka affinities (analogous to magnetic field).
The circulation of the $n$-th tile is 
\beq
\mathcal{B}_n \ \ = \ \ 2 \phi_n
\eeq
If all the circulations are zero, a gauge transformation can be used in order to show 
that $\bm{W}$ is similar to a symmetric matrix $\bm{H}$, hence all the eigenvalues are real, 
as for hermitian Hamiltonians. 
Otherwise we are dealing with the physics of non-hermitian matrices, 
where the spectrum might become complex 
~\cite{Berry2004,bender2007making,Hatano1996,Hatano1997,Shnerb1998,Feinberg1999,KAFRI2004,KAFRI2005,Amir2016,HurowitzCohen2016,HurowitzCohen2016a,Gong2018}

\subsection{Model parameters}
\label{sec:par}
{The motion of the Janus particles is regarded as an hopping process between 
sites of a network, see \Fig{f1}(b). The unbiased transition rates are~$w_b$.  
In the continuum limit we get diffusion with coefficient~$\mathcal{D} \propto w_b$.  
We assume for simplicity that all the~$w_b$ are the same. Some further remarks 
about the implication of having different rates for the vertical flips are presented 
in \App{sec:ndspectrum}.}

The motion in \Fig{f1}(a) is controlled by a non-zero average drift field~$\bar{f}$
and a non-zero average self-propulsion~$\bar{\phi}$. 
Disorder may arise due to the non-homogeneity of the background environment, 
or due to the non-homogeneity of the illumination source. 
Respectively, we distinguish  between   
{\em conservative stochastic disorder} (CSD)  
for which the $f_n$ acquire a random term ${\in [-\sigma_{f},\sigma_{f}]}$, 
and {\em topological stochastic disorder} (TSD)  
for which the $\phi_n$ acquire a random term ${\in [-\sigma_{\phi},\sigma_{\phi}]}$.   
Accordingly the model parameters are $(\bar{f},\bar{\phi},\sigma_{f},\sigma_{\phi})$ 
and the length of the sample~($N$).

\subsection{Types of disorder}
\label{sec:background}

It is customary to distinguish between two types of disorder, 
so called Type-I and Type II \cite{Dyson1953,Ziman1982}. We explain these terms 
in the present context, and highlight new type of disorder that 
has not been illuminated so far. The different types of disorder 
are summarized in Table~\ref{t1}.

Random $f_n$, as in the Sinai model (aka random walk in random environment) \cite{Sinai1983,Derrida1983,BOUCHAUD1990,BOUCHAUD1990a},   
translates, under gauge-like transformation, into Type-I disorder \cite{Shnerb1998}, 
which is a diagonal on-site disorder as in the Anderson model (electron in a random potential).
We shall call it below {\em conservative stochastic disorder} (CSD).

Random $w_b$, as in random resistor network models \cite{Alexander1981}, 
or as in the Debye model (balls connected by non-identical springs), 
translates into Type-II disorder, 
which is an off-diagonal bond disorder.
The latter type of disorder can lead to a percolation-like transition that 
affects the relaxation modes \cite{HurowitzCohen2016}, 
and we shall not consider it further.

In the present work we consider a new type of disorder that we call {\em topological stochastic disorder} (TSD).
This type of disorder originates from having random $\phi_n$, 
and unlike CSD, cannot be gauged-away. 
We note that physically, TSD arises naturally 
also in situations other than active particles.
For example the affinities $\mathcal{B}_n$ may reflect non-conservative
drift-fields that are induced by electro-motive-forces.

With the substitution ${\phi_n \mapsto i \phi_n}$ our TSD
becomes magnetic-field-disorder (MFD) that has been discussed in the past, 
e.g. \cite{Chan1996}  and references therein. 
One should be aware that there is an essential difference between TSD 
and MFD:  the latter has qualitatively the same effect 
as the usual Type-I Anderson disorder, while TSD makes the spectrum complex.

To the best of our knowledge, the only other type 
of non-hermitian disorder that has been 
discussed in the literature, is a random sign disorder (RSD).
See \cite{Amir2016} and references therein.  
It concerns biological networks, where the $w_b$ have random sign, 
corresponding to random excitatory and inhibitory connections.  
It should be realized that RSD has nothing to do with topology: 
the model of \cite{Amir2016} is a single channel tight binding model 
with near-neighbor transitions. 
In contrast, TSD requires at least two channels, 
as in the case of a random magnetic field.  
Also it should be realized that the strength of TSD,
unlike that of RSD, is tunable.

\section{Different routes to complexity}
\label{sec:routes}

We first remind the reader the simplest result for the relaxation spectrum 
of particles that diffuse in a single-channel biased ring. 
This result illustrates the {\em delocalization route to complexity}.
For any non-zero bias the spectrum becomes fully complex ${\lambda = \mathcal{D}k^2 + ivk}$, 
where~$k$ is the wave-number, $\mathcal{D}$ is the diffusion coefficient, and $v$ is the drift velocity. 
Similar expression applies for a tight binding model, see e.g. \cite{HurowitzCohen2016a}.

{The spectrum of a one-channel disordered ring, in the absence of bias, is real.
With added bias some eigenvalues become complex.
The complexity appears only if the bias exceeds a finite threshold, 
aka delocalization threshold \cite{Hatano1997,Hatano1996}.
The eigenstates that correspond to real eigenvalues are then localized,
while those that are associated with complex eigenvalues are extended.}
As discussed in \cite{HurowitzCohen2016a} 
the low relaxation modes (small~$\re[\lambda]$) get delocalized first, while the 
high lying relaxation modes remain real.
 
We consider a two-channel ring of $N$ unit cells with periodic boundary conditions (\Fig{f1}(b)).
This is the simplest example for an active network, 
and we are going to find two additional routes to complexity that have to do 
with the non-trivial topology of the model. 
We distinguish between the circulation that is induced by the drift, 
and the circulation that reflects the self-propulsion, namely,  
\beq
\sum f_n  \ &\equiv& \ N\bar{f} \\
\sum \mathcal{B}_n \ &\equiv& \ 2 N\bar{\phi} 
\eeq
where $\mathcal{B}_n = 2 \phi_n$ is the affinity of the $n$-th tile.
For simplicity we assume that all the couplings are identical ($w_b=1$ for any $b$).
In the absence of disorder, $\bm{W}$ is translationally symmetric 
and can be written very simply using momentum and spin operators:
\beq
\label{eq:w-clean-system}
\bm{W}=  \left(\bm{\sigma}_x - 1\right) 
+ \sum_{\pm} e^{ \pm \frac{1}{2}\left( \bar{f} + \bar{\phi} \bm{\sigma}_z \right) } \left(e^{ \mp i\bm{p}}-1\right)  
\eeq 
The Pauli operator $\bm{\sigma}_x$ term induces the random change in the propulsion direction (vertical transitions in \Fig{f1}),
the $\pm$ terms generate the forward and backward transitions, while the ``-1'' terms provide 
the diagonal elements (``decay rates") that are required for conservation of probability. 
The operators $e^{\pm i \bm p}$ and ${\bm \sigma}_i$ are written using explicit Dirac notations in \App{sec:matrix-elements}.

\begin{figure}[b]
\centering
\includegraphics[]{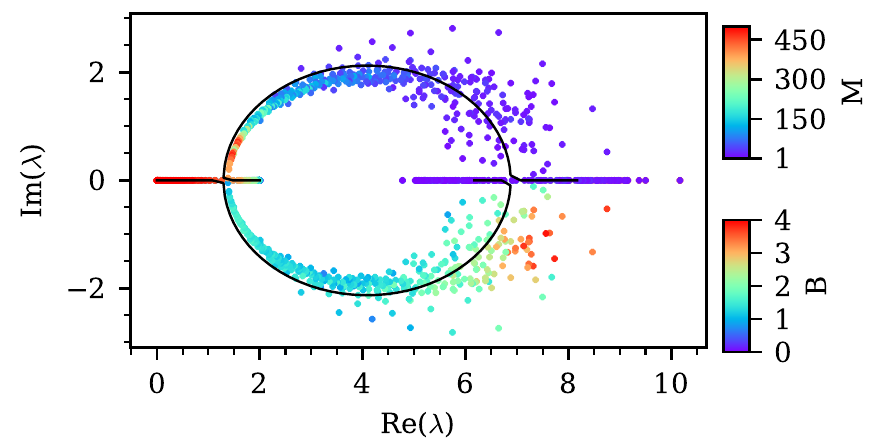} 
\caption{{\bf Representative relaxation spectrum.} 
The spectrum for finite propulsion ($\bar{\phi}=2$) and TSD ($\sigma_{\phi}=1$) with $n=500$.
The eigenvalues are presented in the complex plane.
Each associated eigen-mode is characterized by various measures:  
The real and ${\im[\lambda]>0}$ points are color-coded by the participation number~$M$,
while the conjugate ${\im[\lambda]<0}$ points are color-coded by the effective propulsion~$B$. 
See text for definitions.
The solid line illustrates the spectrum of the non-disordered system \Eq{eq:lambda-prop}.
}\label{f2}
%
\ \\ \ \\
%
\centering
\includegraphics[]{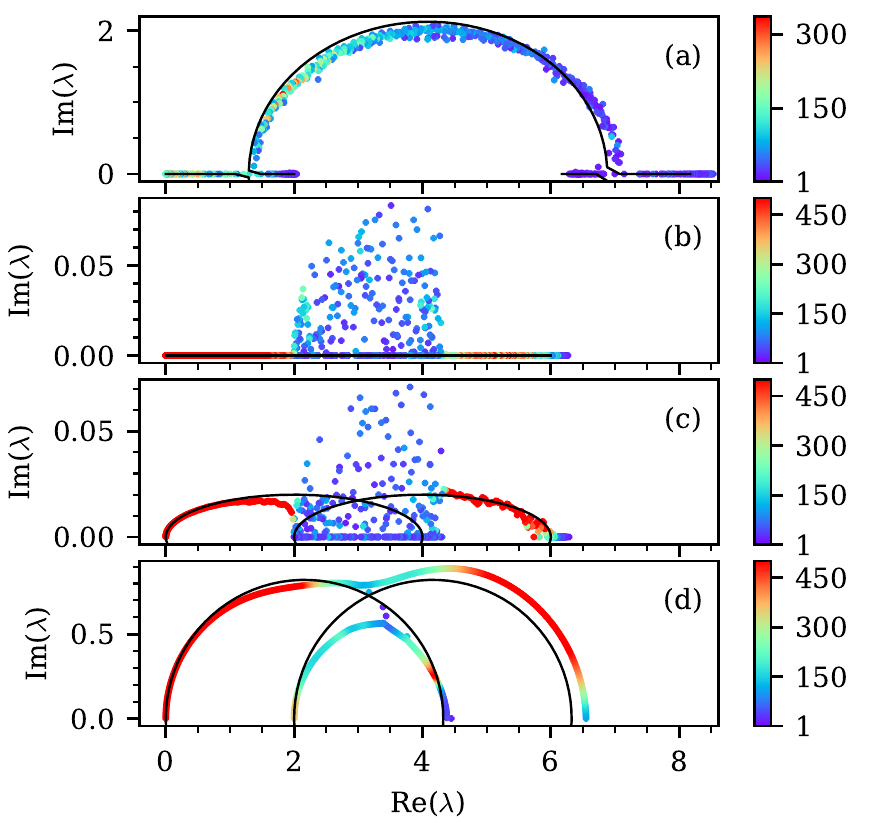} 
\caption{
{\bf Representative relaxation spectra.} 
Same presentation as in \Fig{f2}; conjugate eigenvalues excluded.
{\bf (a)}~Same propulsion as in \Fig{f2}, but with CSD ($\sigma_{f}=1$) instead of TSD.
{\bf (b)}~Same TSD as \Fig{f2}, but the average propulsion is zero. 
{\bf (c)}~Same as (b), but with weak drift ($\bar{f}=0.02$). 
{\bf (d)}~Same as (b), but with stronger drift ($\bar{f}=0.8$). 
}\label{f3}
\end{figure}

We are now in position to explain how each of the parameters 
of the model $(\bar{f},\bar{\phi},\sigma_{f},\sigma_{\phi})$
affects the complexity of the spectrum.

{\em CSD.-- }
Conservative stochastic disorder arises if all the $\phi_n$ are zero 
while the ${f_n \in [-\sigma_{f},\sigma_{f}]}$ have finite dispersion and zero average. 
Such type disorder can be derived from a stochastic potential that features 
activation barriers, as discussed by Sinai and followers \cite{Sinai1983}.
The asymmetry of the $\bm{W}$ matrix can be gauged away, hence it is similar 
to an hermitian matrix, and the relaxation spectrum comes out real.
The corresponding eigenstates are Anderson-localized.

{\em Propulsion.-- } 
Without disorder the relaxation spectrum can be found analytically (see \Sec{sec:symmetry-breaking}).
Adding self propulsion~$\bar{\phi}$, unlike drift, leads to a very different route to complexity, 
that is not related to delocalization of the eigenstates. 
For low self-propulsion the spectrum remains real, 
while above some critical value the relaxation modes undergo a symmetry-breaking transition. 
Consequently a circle of complex eigenvalues appears. 
In \Fig{f2} this circle is indicated by a solid line.
If we add weak stochastic disorder the spectrum is blurred, 
as illustrated in \Fig{f2} for TSD, and in \Fig{f3}(a) for CSD.

{\em TSD.-- }
Another route to complexity has to do with mode mixing due to TSD.
Even if the propulsion is zero on the average, 
we still can have finite dispersion ${\phi_n \in [-\sigma_{\phi},\sigma_{\phi}]}$. 
Then the problem becomes non-Hermitian in a very essential way, 
and part of the spectrum becomes complex.  
This is illustrated in \Fig{f3}(b), where we turn off the propulsion for the system of \Fig{f2}, 
but keep the TSD. It should be clear that if we turn off the propulsion 
for the system of \Fig{f3}(a) the complexity vanishes and we get a real spectrum.

{\em Drift.-- }
Without disorder, finite non-zero drift~$\bar{f}$ has the same effect as for a single-mode ring, 
leading to delocalization of the spectrum. 
We demonstrate in \Fig{f3}(c,d) the drift-induced delocalization route to complexity.
The drift can delocalize the lower (small $\re[\lambda]$), and possibly also the upper (large $\re[\lambda]$)
part of the spectrum, where we have single mode physics. 
We also see the interplay of the drift and the TSD in the middle part of the spectrum 
where the two channels overlap. For strong drift the TSD induces  
an avoided crossing, while for weak drift the TSD-induced complexity predominates.

\section{Topological symmetry breaking}
\label{sec:symmetry-breaking}

The $\bm{W}$ matrix formally operates above an Hilbert space of states 
whose standard representation is 
\beq
\ket{\psi} \ \ = \ \ \sum_{n,s} \psi_{n,s} \ket{n,s} 
\eeq
The right eigenvectors of $\bm{W}$ are the relaxation modes of the network.
The eigenvector that corresponds to the zeroth eigenvalue $\lambda_0=0$ 
is the non-equilibrium steady state (NESS), while all the 
the other eigenvalues are written as $\{ -\lambda_r \}$, with $\re[\lambda_r]>0$.

For our geometry, beside the NESS, there is another special mode with the eigenvalue $\lambda{=}2$.
This can be seen by considering the left eigenvector 
${\ket{\tilde{2}} = \sum_{n} \left( \ket{n,\uparrow} - \ket{n,\downarrow} \right)}$.
All the ${\lambda\neq 2}$  eigen-modes 
are orthogonal to this special left eigenvector, 
hence the sum ${\sum_{n} \psi_{n,s}}$
has to be equal for both polarizations. 
Consequently it is implied that the NESS has equal weight for clockwise 
and anticlockwise motion, while for all the relaxation modes  
the sum of amplitudes vanishes for each direction. 
The same considerations also give the time-dependence of the total polarization
\be{D}
D \ = \  \sum_n D_n \ = \ \sum_n  \left( p_{n,\uparrow} - p_{n,\downarrow} \right)
\eeq
Multiplying \Eq{e1} from the left with $\ket{\tilde{2}}$, 
one obtains a universal decay law ${\dot{D} = -2D}$.

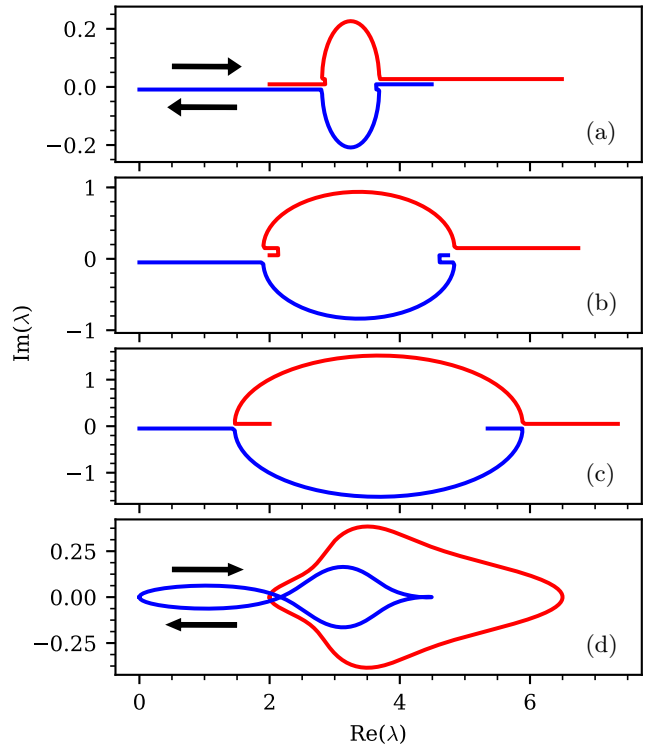
\begin{figure}
\centering
\begin{overpic}{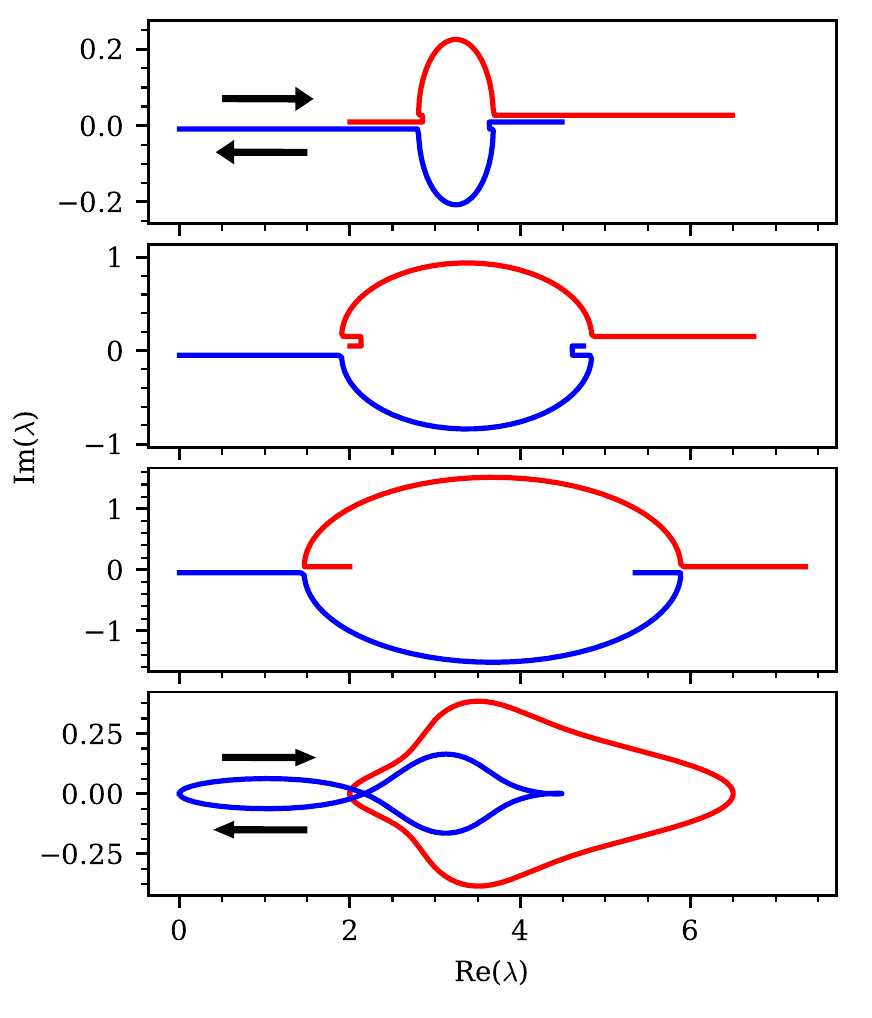} 
\put(75,81){(a)}
\put(75,59){(b)}
\put(75,37){(c)}
\put(75,15){(d)}
\end{overpic}
\caption{\label{fig:prop-only} 
The Bloch spectrum $\lambda_{k,s}$.
The two bands are ${q=1}$ (blue,lower half-plane) 
and ${q=2}$ (red, upper half-plane). 
In panels~(a) to~(c) we have ${\phi=0.98, 1.2, 1.6}$, while ${f=0}$. 
Note that ${\phi_c \approx 0.96}$.  
\Eq{eq:lambda-prop} has been used.   
For presentation purpose the horizontal pieces of the band 
have been shifted off the real axis. 
Panel~(d) is for the same~$\phi$ as in~(a), 
but with an added ${f=0.01}$ that lifts the~$\pm k$ degeneracy.  
The black arrows show the direction in which $\lambda_k$ is changing 
as~${k\in[0,2\pi]}$ is increased from zero, and eventually comes back.   
}
\end{figure}

In the absence of disorder the $\bm{W}$ matrix is block-diagonal 
in the basis $\ket{k,s}$, where $k$ is the wave-number.
For $\bar{\phi}=\bar{f}=0$ the spectrum consists of two bands along the real axis, 
namely, ${\lambda_{k,+} = 2-2\cos(k)}$ and ${\lambda_{k,-} = 4-2\cos(k)}$. 
The existence of the 2-channel topology is reflected by having an overlap ${2<\lambda<4}$.
Note also that all the eigenvalues are doubly degenerate due to ${k\mapsto -k}$ symmetry. 
This holds true also for $\bar{\phi}\ne0$ (we still keep $\bar{f}=0$). 
But now the spectrum becomes complex. The $k$-th block of the $\bm{W}$ matrix is 
\beq
\label{eq:Wk-matrix} 
\bm{W}^{(k)} =  b \bm{\sigma}_x  - ia \bm{\sigma}_z  + c \bm{1} 
\eeq
where $b=1$, and  
\beq
\label{eq:a(k)}
a  &=&  \left[ 2 \sinh\left( \frac{\bar{\phi}}{2} \right) \right] \sin(k)  \\
c  &=&  \left[ 2 \cosh\left( \frac{\bar{\phi}}{2} \right) \right] \cos(k) 
- \left[ 1 + 2 \cosh\left(\frac{\bar{\phi}}{2}\right) \right]
\eeq
The matrix above is similar to a real matrix ${ b \bm{\sigma}_z  - ia \bm{\sigma}_y }$. 
Such matrices are usually encountered if there is an anti-unitary symmetry such as
``T" or ``PT" \cite{Berry2004,bender2007making}.
However we shall stick with the physical representation of \Eq{eq:Wk-matrix}. 
The eigenvalues are  either real or come in complex-conjugated pairs,
namely,
\begin{align}\label{eq:lambda-prop}
\lambda_{k,q} = -\left[ c \pm \sqrt{b^2-a^2}\, \right]
\end{align}
where $q=1,2$ labels the lower and upper band respectively.
The spectrum is real for ${|a|<|b|}$ and complex for ${|a|>|b|}$.
The latter possibility is realized  
if ${2 \sinh(\bar{\phi}/2) > 1}$, 
leading to a critical value for self-propulsion:
\be{14}
\phi_c \ \ \approx \ \ 0.96.
\eeq
\Fig{f2} shows a representative spectrum for $\bar{\phi} > \phi_c$, 
where the solid line is based on \Eq{eq:lambda-prop}.

The eigen-modes are labeled as $\ket{k,q}$. 
In the Bloch sphere representation they reside in the XY or in the YZ plane, 
depending on whether they are associated with real or complex eigenvalues, respectively.     
Close to the so-called exceptional point (${a=b}$) they coalesce into the same Y direction.
Disregarding normalization, the eigen-modes are 
\beq\label{eq:k-pm-propulsion}
\ket{k,q} = \sum_{n} e^{ikn} \left(\ket{n,\uparrow} \ \ \pm \ \ e^{\pm i\varphi} \ket{n,\downarrow} \right) 
\eeq
where ${\tan(\varphi)=a/\sqrt{b^2-a^2}}$.  
Strong self-propulsion (${a>b}$) implies $\varphi=(\pi/2)+i\theta$ with real $\theta$.
Consequently the symmetry with respect to the orientation-direction is broken, 
and the modes become polarized, meaning that   
clockwise modes are separated from anti-clockwise modes. 
On top we note that the $\ket{k,s}$ have a systematic 
degeneracy for ${k\mapsto -k}$.

The spectrum of the non-disordered model is further analyzed 
in \App{sec:ndspectrum}, and is illustrated in a few 
representative cases in \Fig{fig:prop-only}. 
It is composed of two bands. 
As discussed above, in the absence of~$f$  
the bands are deformed into the complex plane provided ${\phi > \phi_c}$. 
Panels (a-c) illustrate this deformation for increasing values of~$\phi$. 
It is important to notice that the $\pm k$ symmetry is not broken, 
hence each of the two $-\lambda_{k,q}$ trajectories 
is degenerated and encloses a zero area. 
This is no longer true of if we add a non-zero~$f$. 
In the latter case the $\pm k$ degeneracy is removed, 
and the $\lambda_{k,s}$ trajectories encloses a finite area.

\section{The introduction of TSD}
\label{sec:TSD}

\begin{figure}[t]
\centering
\includegraphics[]{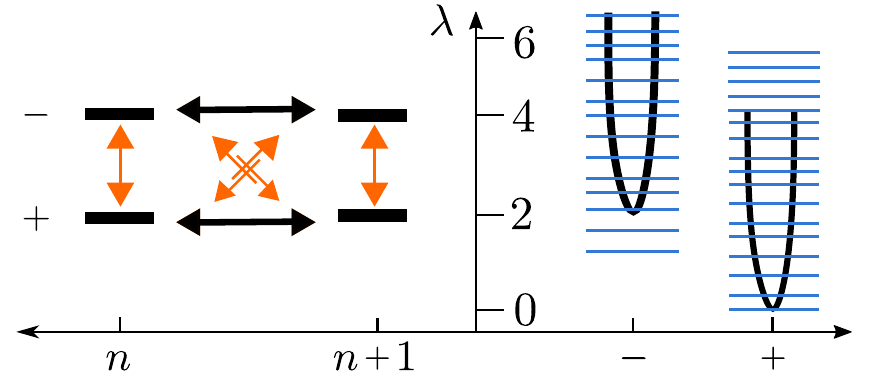} 
\caption{
{\bf Diagonalization procedure.} 
We go from the site-representation $\ket{n,s}$ of \Fig{f1}(b) 
to the mode representation $\ket{\alpha,\pm}$, 
which is illustrated on the left. The diagonal arrows represent anti-hermitian couplings due 
to self-propulsion. The thick double-sided arrows represent the hermitian hopping elements between cells. 
The disorder affects all those couplings,  and also adds vertical hopping elements. 
The unperturbed diagonal `energies' are ${\lambda=2}$ and ${\lambda=4}$. 
With hopping we get two bands ${[0,4]}$ and ${[2,6]}$ that are illustrated by solid line on the right. 
Neglecting the inter-band couplings the spectrum is still real, 
represented by the blue vertical segments. The complex spectrum appears due to band mixing, 
as explained in the main text.
}\label{fd}
\end{figure}

We now consider what happens if the illumination is non-uniform. Thus we have TSD 
with some variation $\sigma_{\phi}$ on top of the average value $\bar{\phi}$.  
At this point one may wonder whether it is feasible to introduce TSD with {\em zero} average propulsion 
(the illuminated particles in \Fig{f1} are always self-propelled in the direction that they are facing). 
After little reflection one realizes 
that it is possible to introduce such TSD if the black-white coating of the particle is reversed 
in its lower half. Then one can use two sources of illumination: upper illumination source that induces 
self-propulsion (for a given polarization) to 
{one direction, and a lower illumination source that induces self-propulsion (for the same polarization) to the other direction.} 
If the two sources are of equal average intensity, the combined effect is to have {\em zero} average propulsion, 
and hence unbiased TSD.

Let us see how the diagonalization procedure for $\bm{W}$ is affected in the presence of non-uniform illumination, 
without assuming any restrictions on the values of $\bar{\phi}$ and $\sigma_{\phi}$.    
The standard site-basis is $\ket{n,s}$. In order to get rid of the vertical coupling we can 
switch to the basis $\ket{n,\pm}$, where $\pm$ are the modes that are defined by $\bm{\sigma}_x$. 
In the absence of propulsion (or TSD) we get two non-interacting chains, see illustration in \Fig{fd}. 
Each chain can be diagonalized hence we go to the basis $\ket{k,\pm}$. 
If we introduce disorder and neglect the inter-band couplings, the spectrum 
is still real, and can be labeled $\ket{\alpha,\pm}$. The $\alpha$-states unlike the $k$-states 
are not `free waves' and become localized as disorder is increased. 
In the  $\ket{\alpha,\pm}$ basis we can write 
\begin{align}
\bm{W}=\bm{H}+\bm{A}, \label{eq:W=HA}
\end{align}
where $\bm{H}$ is hermitian, and $\bm{A}$ is an anti-hermitian matrix due to the self-propulsion.
The disorder-induced hermitian and anti-hermitian couplings are represented, respectively, 
by the vertical and diagonal arrows in \Fig{fd}.

In the absence of disorder $\bm{A}$ couples only states with the same~$k$, 
hence $\bm{W}$ takes the block-diagonal form \Eq{eq:Wk-matrix}
where $\bm{A}=-ia \bm{\sigma}_z$ are the anti-hermitian inter-band couplings.  
Then we get the Bloch eigenstates ${\ket{k,q}}$ where ${q=1,2}$.     
With weak TSD the matrix $\bm{W}$ is no longer block-diagonal. 
It does not require strong disorder in order to induce band mixing. 
The condition for band mixing is to have $\bm{A}$-couplings 
that are larger compared with the level-spacing. 
This is a very easy condition which is implied by perturbation theory, see \App{sec:lalgebra}. 
Consequently very weak disorder is enough 
to induce complexity within the range ${2<\re[\lambda]<4}$.

We note that the appearance of disorder-induced hermitian couplings in $\bm{H}$ of \Eq{eq:W=HA} 
does not change this picture: it scramble the levels of the two bands, but does 
not alter much their density in the overlap region.

\begin{figure} 
\centering
\begin{overpic}{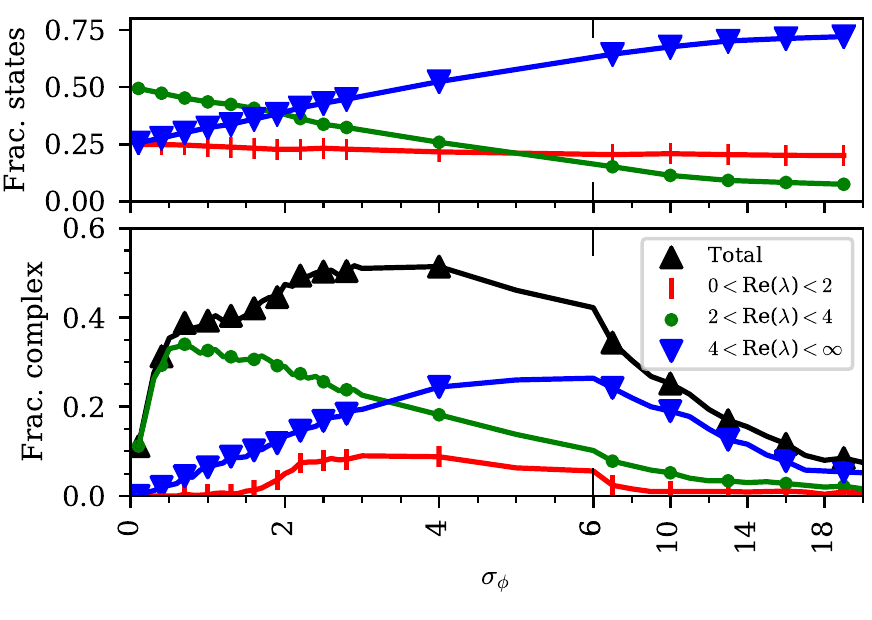} 
\put(16,64.5){(a)}
\put(16,40){(b)}
\end{overpic}
\caption{
{\bf The spectral distribution of the eigenvalues.} 
{\bf (a)} 
The fractions of eigenvalues per spectral window versus 
the disorder strength $\sigma_{\phi}$, 
{while ${\bar{\phi}=\bar{f}=\sigma_f=0}$.}
{\bf (b)}~ 
The fraction of complex eigenvalues (out of all eigenvalues), 
divided into the different spectral windows.
In some realizations of the system there are residual complex eigenvalues 
in the first window (${\re[\lambda]<2}$), with small imaginary part.
It is not clear whether these are numerical issues or not.
The threshold for complexity here is ${\im[\lambda] > 10^{-4}}$.
The plots are for rings of ${N=500}$ cells, 
{with  ${\phi_n = \alpha_n \sigma_{\phi}}$, 
where $\alpha_n \in [-1,1]$ are random numbers.
Each point corresponds to a different $\sigma_{\phi}$, 
while the $\alpha_n$ are not modified.}
}
\label{f5}
%
%
\ \\ \ \\
%
%
\centering
\begin{overpic}{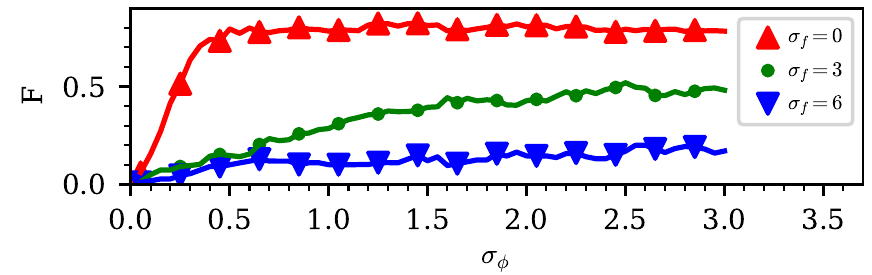} 
\put(16,26){(a)}
\end{overpic}
\begin{overpic} {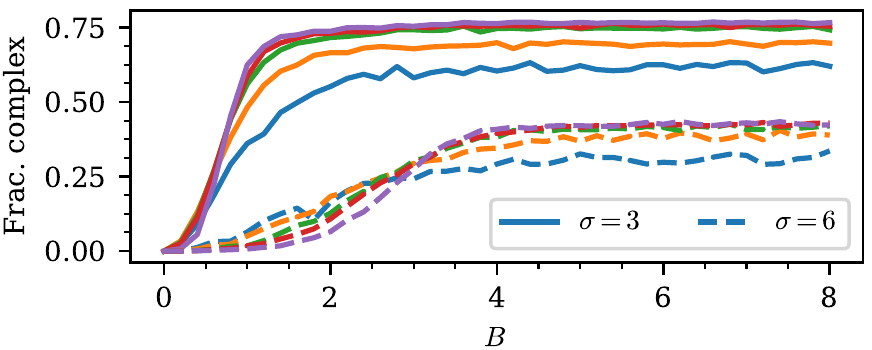}  
\put(16,34){(b)}
\end{overpic}

\caption{
{\bf The fraction of complex eigenvalues.} 
{\bf (a)}~The fraction of complex eigenvalues 
relative to the number of eigenvalues
in the second window ($\re[\lambda] \in [2,4]$),
{for the same ring of \Fig{f5}, 
and for non-zero $\sigma_f$ too.} 
In the latter case the fraction becomes smaller.
{\bf (b)}~The total fraction of complex eigenvalues 
for a simple ring. Each point in the plots is averaged over 100~realizations.
The solid and dashed lines are respectively for $\sigma{=}3$ and $\sigma{=}6$.
The different lines (from bottom to top at ${B>4}$) 
are for rings of length ${L = 5, 10, 20, 30, 60}$. 
}
\label{f5b}
\end{figure}

The explanation above illuminates why uniform $\phi_n$, unlike  
random $\phi_n$ that has the same average intensity, 
requires a finite threshold \Eq{e14} in order to induce complexity in the spectrum.
\Fig{f5} displays how the overall fraction of complex eigenstates  
depends on $\sigma_{\phi}$, while \Fig{f5b}(a) 
shows their percentage within the range ${2<\lambda<4}$.  
{
In order to further explain the consequences of TSD, 
we now turn to characterize the eigenstates.
}

\section{Topological characterization of the eigen-modes}
\label{sec:characterization-eigenmodes}

Assuming square-integrable normalized eigenstates we 
formally define ${P_{n,s} = |\psi_{n,s}|^2 }$ and ${P_n=\sum_s P_{n,s}}$, 
such that ${\sum P_{n} = \sum P_{n,s}=1}$.
Additionally we define, using an harmonic average, 
a topological weight for each tile:
\beq
P_n^* \ \  =  \ \  8 \left[ \sum_{\text{site} \in n}  P_{n,s}^{-1} \right]^{-1}
\eeq
where $\text{site} \in n$ refers to the 4 sites from which the $n$-th tile is formed.
The prefactor is chosen such that ${P_n^* = 1/N}$ for a uniform occupation.
A vanishingly small $P_n^*$ means that the $n$-th cell does not form a closed ring.

It is now possible to  introduce 
several measures that characterize 
a given eigen-modes:
\beq
M \ \ &=& \ \ \left[ \sum_{n,s} P_{n,s}^2 \right]^{-1} \\
L \ \ &=& \ \ \left[ \sum_{n} P_{n}^2 \right]^{-1} \\
L^{*} \ \ &=&\ \ \left[ \sum_{n} P_n^* \right] L\\
Q \ \ &=& \ \ \sum_{(n,s) \in \text{floor}} P_{n,s} \label{eQ} 
\eeq
The first two measures characterize the volume that is occupied 
by the eigen-modes: $M$ is the number of sites that participate 
in the formation of the eigen-mode, while $L$ is the respective 
localization length. The topological localization length $L^*$ 
is further discussed below. The definition and the 
significance of $Q$ will be discussed in the next section.

\begin{figure}
\centering
\begin{overpic}{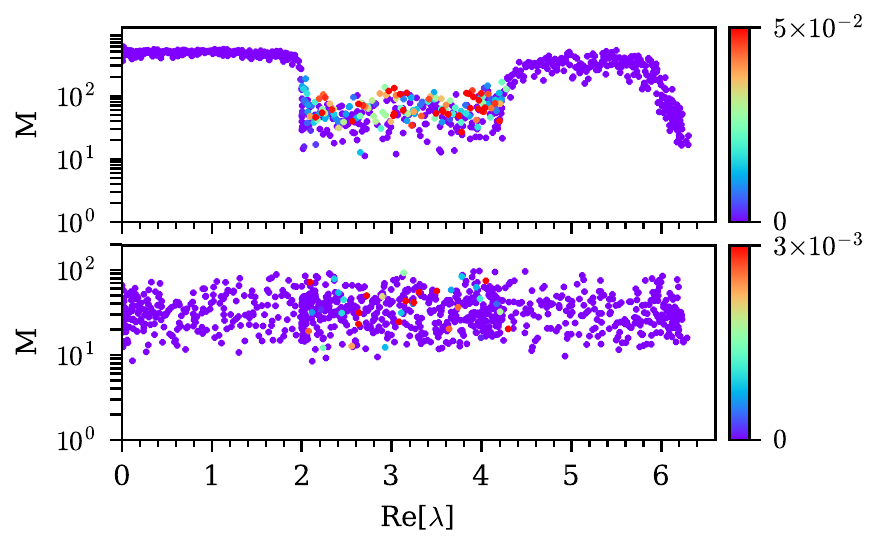} 
\put(16,40){(a)}
\put(16,15){(b)}
\end{overpic}
\begin{overpic}{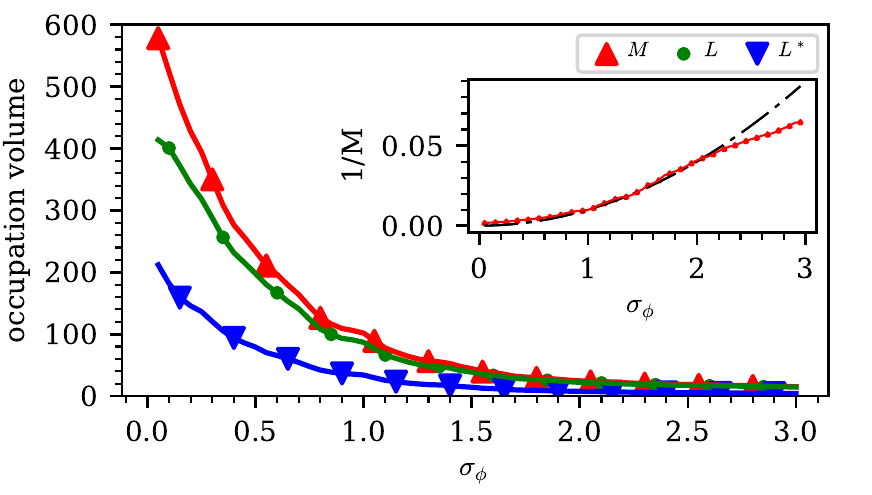} 
\put(16,14){(c)}
\end{overpic}
\caption{{\bf The participating sites for each eigenmode.} 
{\bf (a)}~The participation number $M$ versus $\re[\lambda]$ 
for $\sigma_{\phi} {=} 1$ and $\sigma_f {=} 0.1$, 
{while ${\bar{\phi}=\bar{f}=0}$.} 
{\bf (b)}~Same as (a) but with the disorder 
parameters swapped: $\sigma_{f} {=} 1$ and $\sigma_\phi {=} 0.1$.
{\bf (c)}~The various occupations volumes versus $\sigma_{\phi}$.
In~(a) and~(b) the points are color coded by $\im [\lambda]$. 
In~(c) each point is an average over all the eigenstates 
within ${2<\re[\lambda]<4}$. The inset shows that the inverse localization length 
has roughly quadratic dependence on the disorder strength, as in the Anderson model.  
}\label{f4}
\end{figure}

It is important to realize that the eigenstates might be polarized,
meaning that ${P_{n,\uparrow}-P_{n,\downarrow}}$ is not zero.
Polarization can arise either due to symmetry breaking, as discussed in \Sec{sec:symmetry-breaking},
or due to the formation of a floor-level, which is discussed in the next section.
For a strictly polarized eigenstate ${L=M}$ as opposed to ${L=M/2}$.  
{Numerical results for $M,L,L^*$ are presented in \Fig{f4}}.

The topological localization length $L^*$ reflects 
the effective circulation which is experienced 
by a given eigen-mode. It is determined by 
the total topological weight $\sum P_n^*$, 
which is the occupation probability 
of the region that experiences propulsion. 
If the total topological weight is much smaller than unity, it means 
that the non-hermiticity can be gauged away from the volume that supports the eigenmode, hence the eigenvalue is real (or with very small imaginary part).
If the total topological weight is non negligible, it makes sense 
to define the effective circulation that is experienced by the eigen-mode 
as follows: 
\begin{align}
\label{eq:B}
B \ \  =  \ \ \sum_n P_n \mathcal{B}_n
\end{align}
In the absence of disorder, the eigen-modes are uniform, 
and and we get ${B=2\bar{\phi}}$. In the presence of disorder, 
the eigen-modes get localized, but if they are uniform within 
the localization volume (with zero polarization) 
we still get ${B \approx 2\bar{\phi}}$. 
On the other extreme, if the eigen-modes are completely 
polarized we get a vanishing~$B$. For intermediate 
situation, where the eigen-mode is supported partially by 
topologically connected cells, and partially by dangling sites,
the bare ${2\bar{\phi}}$ is multiplied by the total topological 
weight of the eigen-mode.

\begin{figure}
\centering
\begin{overpic}{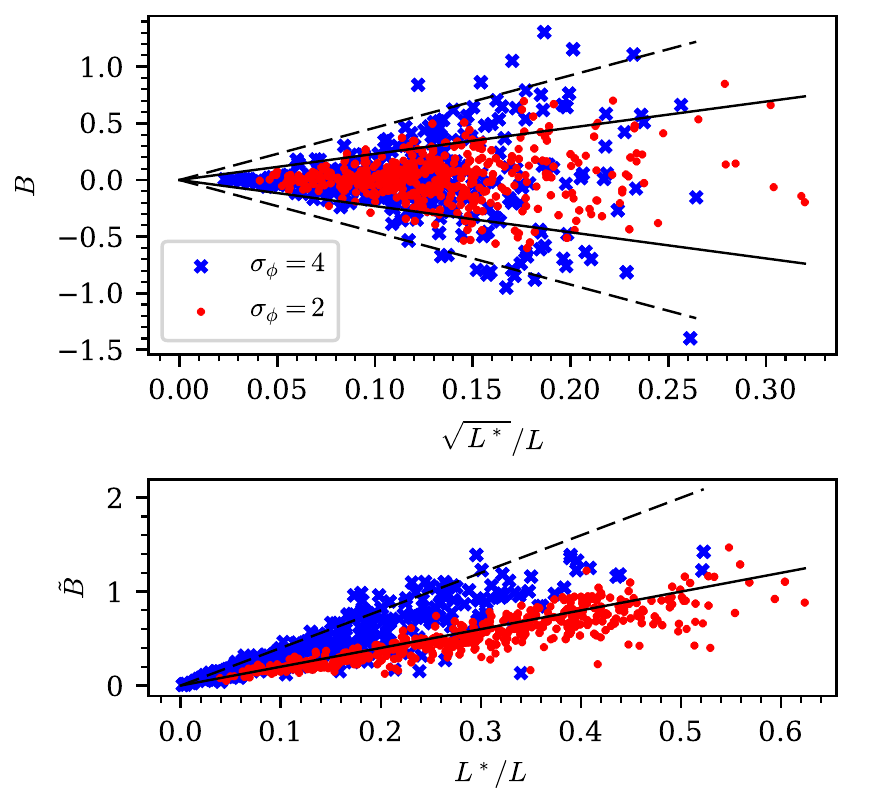} 
\put(19,85){(a)}
\put(19,32){(b)}
\end{overpic}
\caption{\label{fig:Bn-vs-L} 
\textbf{(a)}~
The effective circulation $B$ of the eigenstates
versus their $\sqrt{L^{*}}/L$, 
calculated for TSD with $\sigma_{\phi} = 2$ (red) and $\sigma_{\phi} = 4$ (blue), 
{while ${\bar{\phi}=\bar{f}=\sigma_f=0}$.}
The solid and dashed lines are given by \Eq{e22}.
\textbf{(b)}~
We define ${\tilde{B} =\sum_n P_n |\mathcal{B}_n|}$
and verify that it agrees with the estimate $\sim (L^{*}/L) \sigma_{\phi}$.
Both panels refer to the same set of eigenstates, 
namely, those that reside in the spectral window ${2<\re[\lambda]<4}$.
}
\end{figure}

The effective circulation for each state in \Eq{eq:B} can be estimated
using the measures $L$ and $L^{*}$ as follows:
By definition, ${\sum P_n = L^{*}/L}$.
There are $L^{*}$ terms in the sum, 
accordingly each term can be estimated as $\sim 1/L$.
It follows that $B$ is normally distributed 
with zero mean and standard deviation 
\be{22}
\text{Std}(B) \ = \ \frac{\sqrt{L^{*}}}{L} \text{Std}(\mathcal{B}_n)
\ = \ 2 \frac{(2 \sigma_{\phi})}{\sqrt{12}} \frac{\sqrt{L^{*}}}{L}
\eeq
This estimate is tested in \Fig{fig:Bn-vs-L}.
We conjecture that $B$ affects the complexity of the eigen-mode. 
A hand-waving argument that supports this conjecture goes as follows:
All the asymmetric transition of dangling bonds can be gauged away 
using a similarity transformation; hence $\bm{W}$ is similar to 
a matrix ${H+A}$ where $H$ is real and symmetric, while the anti-symmetric matrix~$A$
is supported only by the topologically connected cells.
Multiplying  ${\bm{W}\psi=\lambda\psi}$ from the left by $\psi\+$ 
we deduce that ${\im[\lambda] = \sum A_{ij} \im[\psi_i^* \psi_j]}$.
Consequently we conclude that from statistical perspective  ${\im[\lambda]}$ 
is proportional to the topological weight of the eigen-mode.

The above conjecture provides a qualitative explanation for the 
remarkable difference between TSD and CSD in \Fig{f2} and \Fig{f3}(b) respectively.
The transverse scattering of the complex eigenvalues in the former case becomes 
larger as the localization volume $M$ becomes smaller.
Modes with larger $B$ experience (by definition) a larger effective propulsion,  
and therefore they are pushed to a larger radius. 
CSD, unlike TSD, does not have a systematic ($M$ dependent) effect on $B$,  
because the $\mathcal{B}_n$ are the same for all cells.

One wonders what determines the complexity saturation value of~$F$ in \Fig{f5b}(a).
For one dimensional rings that were studied in \cite{HurowitzCohen2016}
an analytical treatment has been introduced:  
for stronger disorder the saturation value becomes smaller, 
and the approach to this value is smeared, as illustrated in \Fig{f5b}(b). 
{We now use these results in order to provide 
a simple explanation for the disorder dependence 
of~$F$ in our model system.}
{A given eigenstate with localization length~$L$ lives
on a sublattice of size~$L$. Ignoring the detailed bond structure,  
it can be regarded as an eigenstate of a local-ring of size~${\sim}L$.
The $\sigma_f$ of CSD is by definition the $\sigma$ disorder 
for such local-ring, while the $\sigma_{\phi}$ of TSD 
controls the effective~$B$ and hence analogous to the affinity 
of the one dimensional ring. It is therefore expected 
that $\sigma_f$ rather than $\sigma_{\phi}$ will have 
a dominant influence on~$F$.}
{Indeed we see in \Fig{f5b}(a) that~$F$ is sensitive 
to CSD: increasing $\sigma_f$ unlike increasing $\sigma_{\phi}$ 
affects the saturation value.}
It is true that further increase of $\sigma_{\phi}$ affects the~$L$ 
of the eigenstates too, but this has almost no implication. 
To see why, we illustrate in \Fig{f5b}(b) how the~$F$ of a simple 
ring is affected by its length~$L$, which plays the role of localization length 
in the model under study.

\section{The formation of the floor level}
\label{sec:floor-level}

\begin{figure}
\centering
\begin{overpic}[width=9cm]{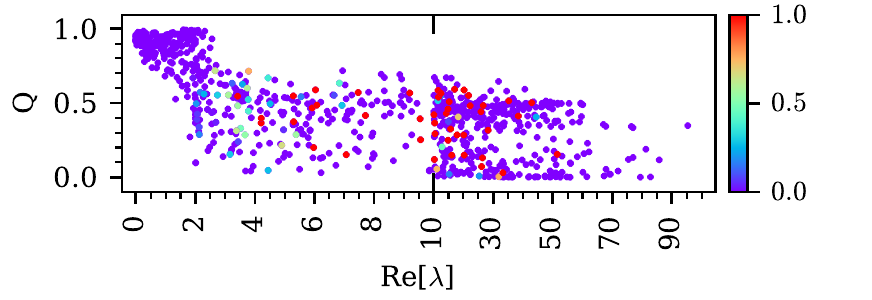} 
\put(15,15){(a)}
\end{overpic}
\begin{overpic}{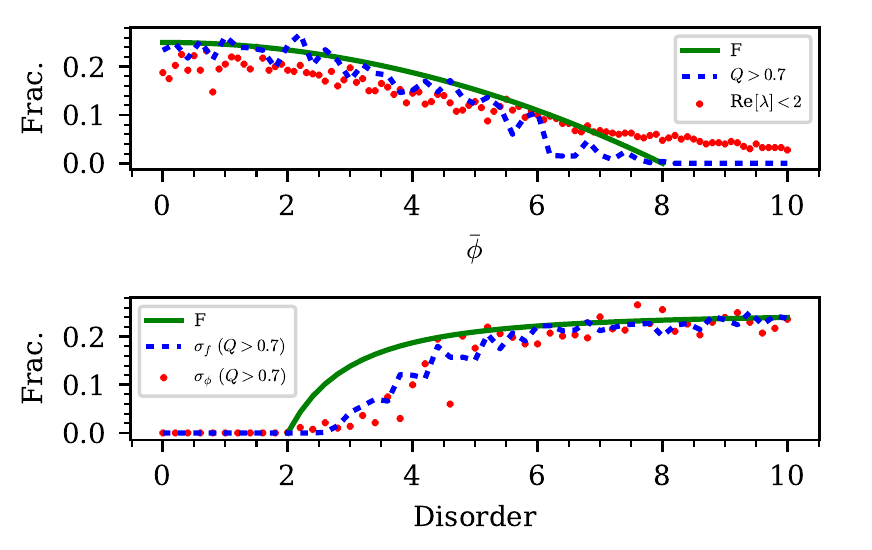} 
\put(16,46){(b)}
\put(86,16){(c)}
\end{overpic}

\caption{{\bf The floor-level occupation.} 
{\bf (a)}~The floor-level occupation $Q$ versus $\re[\lambda]$ 
for $\sigma_{\phi} {=} 8$, {while ${\bar{\phi}=\bar{f}=\sigma_f=0}$.}
{\bf (b)}~The fraction of floor-level eigenstates drops down from 25\% as $\bar{\phi}$ is increased. 
Floor-level eigenstates are defined as those that have ${Q>0.7}$. 
This is compared with $F$ from \Eq{eq:percent-floor-band}, 
and with the number of states with  $\re[\lambda]<2$ ($\sigma_{\phi} = 8$). 
{\bf (c)}~The fraction of floor-level eigenstates versus $\sigma_f$ and $\sigma_{\phi}$. 
The average propulsion is ${\bar{\phi}=2}$. 
}\label{f6}
\end{figure}

{In the presence of strong disorder, 
the \textit{floor-level} sites are those that serve as {\em sinks} 
for the probability flow. See \Fig{f1}(c) for illustration. 
The NESS and the low lying relaxation modes mainly occupy these floor-level sites. 
This hypothesis is established by \Fig{f6}(a), 
where we plot the floor occupation $Q$ that has been defined in \Eq{eQ}.}

In \Fig{f5}(a) we show how the eigenvalues are distributed with respect to $\re[\lambda]$. 
Generally speaking we see that the spectrum is stretched upward along $\re[\lambda]$.
This can be easily explained noting that the diagonal elements of the $\bm{W}$ matrix 
become very large for strong disorder. Namely,  
\be{20}
\gamma_{n,s} \ \ = \ \  1 + e^{\phi_n/2} + e^{-\phi_{n-1}/2}
\eeq
But a careful look reveals that within ${\re[\lambda] \in [0,2]}$ we have an approximate 25\% fraction of real eigenvalues, 
irrespective of the TSD strength. The 25\% is not surprising in the limit of weak disorder:
their reality is implied by the band structure. But their presence and reality persist also for very strong disorder
due to the formation of the floor level. 
From \Eq{e20} it follows that for sink site ${1<\gamma <3}$.
Furthermore, as the disorder is increased  ${\gamma \rightarrow 1}$. The hopping between the floor-level states (via high lying states) leads to the formation of the floor-level band, as established by \Fig{f4}(c). If we have TSD only, the fraction of floor-level sites is 25\%. Adding propulsion this fraction becomes
\beq\label{eq:percent-floor-band}
F_\text{floor} \ = \ \dfrac{(\sigma_{\phi} - \bar{\phi}) (\sigma_{\phi} + \bar{\phi})}{4 \sigma_{\phi}^2}
\eeq
Consequently, as the disorder is increased we expect a crossover from band-structure implied occupation 
to the floor-level implied occupation that is illustrated in \Fig{f6}(a).

Summarizing the TSD case (with zero average propulsion) we realize that in the absence of disorder, 
the low lying eigen-modes are real and non-polarized, because they all belong to a single-channel 
symmetric mode $\ket{k,+}$ of \Eq{eq:k-pm-propulsion} with ${\varphi=0}$.
Increasing the disorder strength, the low lying eigen-modes occupy only the floor level,
hence become polarized (2~sites with the same $n$ cannot both serve as sinks), 
and therefore remain real. 
Note however that we cannot exclude that what we call here ``real"
possesses a very small imaginary part due some residual hopping. On the basis of the numerics 
it is difficult to obtain a conclusive statement, but from practical (physical) point 
of view such conclusive statement is of no importance.

The same calculation of \Eq{eq:percent-floor-band} holds if we have CSD instead of TSD, 
namely, with $\sigma_{\phi}$ replaced by $\sigma_f$. See \Fig{f6}(c). 
In contrast with the TSD case, in the strong CSD limit, 
the sinks come in strongly coupled pairs 
(2~sites with the same $n$ are coupled by a direct vertical bond ${w_{\text{flip}}=1}$,  
rather than via high lying states with ${w_{\text{eff}}\ll 1}$).
Consequently half of the floor states have ${\lambda>2}$ 
as implied by \Fig{fd}. Accordingly, for strong CSD the fraction of states 
within $\re[\lambda] \in [0,2]$ is approximately $12.5\%$ rather than $25\%$.

\section{The topological index}
\label{sec:topo}

The {\em bulk-edge correspondence} principle suggests that localized states should appear 
at interfaces, connecting regions of the sample characterized by a different topological number.
Below we  illuminate the relation between this statement, and the floor-level 
phenomenology that has been introduced in the previous section. 

A translationally invariant sample can be characterized by the winding number  
\beq
w \ = \ \frac{1}{2\pi i} \int_0^{2\pi} dk \frac{d}{dk} \ln\left( \det \left[ W^{(k)} \right] \right),
\eeq
which counts the number of times that the eigenenergies encircle the zero energy. 
Similar to the case of a vanishing band-gap in the Hermitian case, 
the winding number is ill defined when dealing with a conservative system,  
that always has a ${\lambda=0}$ eigenvalue. 
To circumvent this problem, following~\cite{Murugan2017,Vaikuntanathan2017}, 
one has to introduce an $F$-bias, as explained in \App{sec:ndspectrum}.

Considering an interface between two bulk regions "L" and "R" 
the topological index is defined as  
\be{dw}
\delta w \ \ = \ \ w_L-w_R
\eeq
The interface will localize left (right) zero-energy edge-modes, 
if in some finite neighborhood of ${F=0}$, 
the index $\delta w$ is positive (negative) independently of~$F$. 
It is important to realize that two bands are not required
for observing topological phenomena in non-hermitian Hamiltonian, 
which stands in contrast with Hermitian systems~\cite{Gong2018}.

In \App{sec:ndspectrum} we calculate the topological number of a translationally invariant system given by \Eq{eq:w-clean-system}. We find that a non-zero topological index is associated with interfaces between regions with opposite drift field, independent of the self propulsion.
We further observe that probability density can accumulate also at interfaces between regions that have the same topological number. We point out (see last paragraph of \App{sec:ndspectrum}) that the localization in the latter case is less pronounced, and diminishes if the length of the non-disordered regions is increased.

The above observations lead to the conclusion that CSD is more effective (compared with TSD) 
in introducing localized states. The implication of this observation is demonstrated in \Fig{f4}.
We see in panel~(b) a remarkable increased in the likelihood to observe 
eigen-modes with small~$M$.  Another way to phrase this conclusion is to 
say that eigen-modes that reside in the floor level tend to localize if CSD is dominating, 
and tend to be more extended if TSD is dominating.

\section{Summary and discussion}
\label{sec:conclusions}

The relaxation modes of a stochastic network can be either over-damped 
or under-damped depending on whether their $\lambda$-s are real or complex. 
In a non-active unbiased disordered network, say a ring, the relaxation is over-damped. 
But if we add bias (finite $\bar{f}$) the low modes become delocalized and we can have under-damped relaxation, which is associated with correlated currents over the whole ring.

The picture of relaxation is much richer if we consider an active network.
Without disorder the self-propelled motion (finite $\bar{\phi}$) implies 
that above a critical value ($\phi_c$) the relaxation modes become polarized
due to topological symmetry breaking. 
Once disorder is taken into account the picture changes dramatically.
An emergent feature of active networks is a novel type of disorder - TSD. 
Random $\phi_n$, unlike uniform $\phi_n$, does not require a finite threshold
to induce complexity in the spectrum.
{Either way, in the absence of drift, the underdamped modes 
are gaped away from the ${\lambda=0}$ NESS, 
and characterized by a decay rate that is comparable with 
the transition rates (${w_b=1}$ in our numerical examples).}

Thus we have highlighted that a stochastic network 
can exhibit 3 routes to complexity: 
(a)~delocalization of a relaxation mode due to drift;
(b)~topological symmetry breaking of a relaxation mode; 
(c)~TSD-induced band-mixing of real relaxation modes. 
The two latter routes to complexity reflect that we are dealing with an active network.
These mechanisms are local is some sense, 
and are not associated with global delocalization of the eigen-modes.

{Additionally we have shown a new theme that distinguishes  
active matter from passive matter, namely, the emergence of {\em under-damped} 
relaxation modes. Here one should be careful with terminology. 
Under-damped relaxation has been discussed in the context of active matter \cite{Ai2017}, 
but it was associated with the finite inertia of the particles.
In contrast, the type of under-damped relaxation that we have discussed here 
is specifically related to the essence of having an active network, 
namely, the non-conservative nature of the active stochastic field.}

We have presented a detailed investigation of the Fourier-Laplace spectrum 
for a minimal model of an active network \Fig{f1}.  
A time-domain illustration of the dynamics has been displayed in \Fig{f10}
{by means of the local polarization $D_n$}.  
This illustration shows how the under-damped relaxation 
due to self-propulsion is blurred by the introduction of disorder.
Our objective was to provide a quantitative analysis for this dynamical behavior.

In \Sec{sec:characterization-eigenmodes} we have introduced several measures 
for the characterization of a relaxation mode: 
the number of lattice sites that support the mode~($M$);  
the number of floor-level sites that are involved~($Q$); 
their localization lengths~($L$);  
and the effective circulation that they experience~($B$).
These measures helped to gain a deeper insight into the spectrum. 

A few remarks are in order: 
{\bf (1)}~ 
If the average self-propulsion is zero, the effect of CSD is to stabilize the 
reality of the spectrum, while TSD induces complexity in the central part 
of the spectrum via band-mixing. 
{\bf (2)}~
If we further increase the stochastic disorder 
the fraction of complex eigen-modes become smaller. There are two issues here:
the formation of the floor-level due to the effective dilution of the lattice; 
and the fragmentation of the lattice into smaller regions that support 
the localized eigen-modes. 
{\bf (3)}~
Opposing to the common perspective that ties between
delocalization and complexity in a single channel system \cite{Hatano1996}, 
TSD both makes the spectrum complex and localizes the states.
%
{\bf (4)}~
The effect of TSD can be distinguished from the effect of CSD
also if the average self-propulsion is not zero (finite $\bar{\phi}$). 
The CSD affects democratically all the under-damped modes,  
while the TSD has larger effect on the more localized modes.

We can adopt a more general perspective with regard to the floor-level phenomenology, that can be applied for any active network. Strong stochastic disorder, irrespective of its nature, induces {\em lattice dilution}, leading to the formation of a floor-level that is spanned by the sites that serve as sinks for the stochastic flow. Consequently the effective dimensionality of the lattice reduces. In our geometry the floor-level sites form a single channel chain, hence a robust reality is gained within the floor-level band.

The floor level consists of local sinks of the stochastic flow. In particular local sinks appear at interfaces between segments characterized by a different {\em topological numbers} that are determined by the sign of the drift flow, irrespective of the self-propulsion. Still, we observe that probability density can accumulate at interfaces between regions that have the same topological number, e.g. in the presence of TSD and uniform drift. Our numerical analysis shows that the degree of localization in the latter case is less pronounced and smeared away if the non-disordered regions are lengthy. This observation highlights the role of topological protection and its implication on localization in non-equilibrium stochastic flow.

\appendix

\section{The $W$ matrix}
\label{sec:matrix-elements}

The matrix $\bm{W}$ can be regarded as the representation of 
a non-hermitian Hamiltonian. It consist of 3~terms:
\beq \nonumber
\bm{W} \ = \ \bm{W}_{\text{flip}} + \bm{W}_{\text{hop}} - \sum_{n,s} \ket{n,s} \gamma_{n,s} \bra{n,s}
\eeq
Using the Dirac's bra-ket notations, 
the explicit expressions for the flipping and hopping terms are 
\beq \nonumber 
\bm{W}_{\text{flip}} &=& \sum_{n}
\Ketbra{n,\uparrow}{n,\downarrow} + \Ketbra{n,\downarrow}{n,\uparrow} 
\\ \nonumber 
\bm{W}_{\text{hop}} &=& \sum_{n,s}
\Ketbra{n{+}1,s}{n,s} e^{\frac{\mathcal{E}_{n,s}}{2}}
+ \Ketbra{n,s}{n{+}1,s} e^{-\frac{\mathcal{E}_{n,s}}{2}} 
\eeq
with $\mathcal{E}_{n,s}$ that are given by \Eq{eq:epsilon-equals-topo-field}.
The decay rate are implied by conservation of probability:
\beq
\gamma_{n,s} \ = \ 1 + e^{\mathcal{E}_{n,s}/2} + e^{-\mathcal{E}_{n-1,s}/2} 
\eeq
The translation operators $e^{\pm i \bm p}$ of \Eq{eq:w-clean-system} 
are defined by $e^{\mp i \bm p} \ket{n,s} = \ket{n{\pm}1,s}$.
The ${\bm \sigma}_i$ operators are defined by $\BraKet{n',s'}{\bm{\sigma}_i}{n,s} = (\sigma_i)_{s's} \delta_{n'n}$ 
in terms of Pauli matrices.

\section{The non-disordered spectrum}
\label{sec:ndspectrum}

Pedagogically it is useful to consider a single-channel 
tight binding model, which is biased by stochastic field~$f$. 
Additionally we introduce an $F$-bias~\cite{Murugan2017,Vaikuntanathan2017}, 
which affects the off-diagonal rates, but not the diagonal elements. 
Accordingly 
\beq
\label{eq:w-schannel}
\bm{W}=  
- 2\cosh\left(\frac{f}{2}\right) 
+ \sum_{\pm} e^{ \pm \left(\frac{f}{2}+F\right) } \ e^{ \mp i\bm{p}}
\eeq 
The Bloch spectrum is $\{-\lambda_k \}$ with 
\beq \nonumber
\lambda_k = 2\cosh\left(\frac{f}{2}\right) 
&-& 2\cosh\left(\frac{f}{2}{+}F\right)\cos(k) 
\\ \nonumber
&+& i2\sinh\left(\frac{f}{2}{+}F\right)\sin(k)  
\eeq
This spectrum goes through the origin for ${F=0}$, 
which reflects the existence of the NESS for 
a conservative matrix. But for any non-zero $F$ 
we can define the winding number~$w$ of the 
$-\lambda_k$ trajectory relative to the origin. Namely, 
\be{wchain}
w \ = \ \text{sign}\left[ |f+2F|- f \right] 
\eeq 
If we have two regions (left and right) that do not have the same~$f$, 
the difference ${\delta w \equiv w_L-w_R}$ is well defined 
in the limit ${F\rightarrow 0}$ and does not depend whether we 
take the limit from the positive or from the negative side. 
For the topological index  of \Eq{edw} we get ${\delta w =1}$. 
By the  {\em bulk-edge correspondence} principle it is implied 
that  a bound state should reside at one of the two interfaces between the two 
bulk regions which acts as  a sink for the flow. 
(We assume periodic boundary conditions, so the interface is in fact two locations along the ring.) 

We now can consider on equal footing our model system \Eq{eq:w-clean-system},  
{where we have two coupled chains, with flip rate ${b=1}$. 
For generality we assume below general~$b$.} 
Here the winding number is calculated from the ${2\times2}$ matrix $\bm{W}^{(k)}$. 
As in the single-channel example the topological-index calculation requires to 
introduce an $F$-bias. The resulting matrix reads:
\beq
\nonumber
\bm{W}^{(k)} 
&=& b\sigma_x + 2\cosh\left(\frac{f+\phi\sigma_z}{2}+F-ik\right) \\
&& -b-2\cosh\left(\frac{f+\phi\sigma_z}{2}\right)
\eeq
with eigenvalues $\{-\lambda_k \}$, where
\beq
\nonumber
&&\lambda_{k,q} = b-4\cosh\left(\frac{\phi}{2}\right)\sinh\left(\frac{f{+}F{-}ik}{2}\right)\sinh\left(\frac{F{-}ik}{2}\right)
\\
\nonumber
&&\mp \sqrt{b^2+16\sinh\left(\frac{\phi}{2}\right)^2\cosh\left(\frac{f{+}F{-}ik}{2}\right)^2\sinh\left(\frac{F{-}ik}{2}\right)^2}
\eeq
The index $q=1,2$ corresponds to $\mp$. Note consistency with \Eq{eq:w-schannel} 
upon the substitution ${b=\phi=0}$. 
{We get two partially overlapping bands provided ${b<2}$.}
The spectrum in a few representative cases has been illustrated in \Fig{fig:prop-only}. 
The $\pm k$ degeneracy is removed if we add a non-zero~$f$, 
which parallels the single-channel analysis.
This has been demonstrated in \Fig{fig:prop-only}(c).   
With an additional $F$-bias the loop looks similar but does not go through the origin.
Our analysis shows that the presence of a finite self propulsion 
does not alter the topological index. 
Namely, the expression for~$w$ is \Eq{ewchain} as for a single chain.

\begin{figure}
\centering
\begin{overpic}{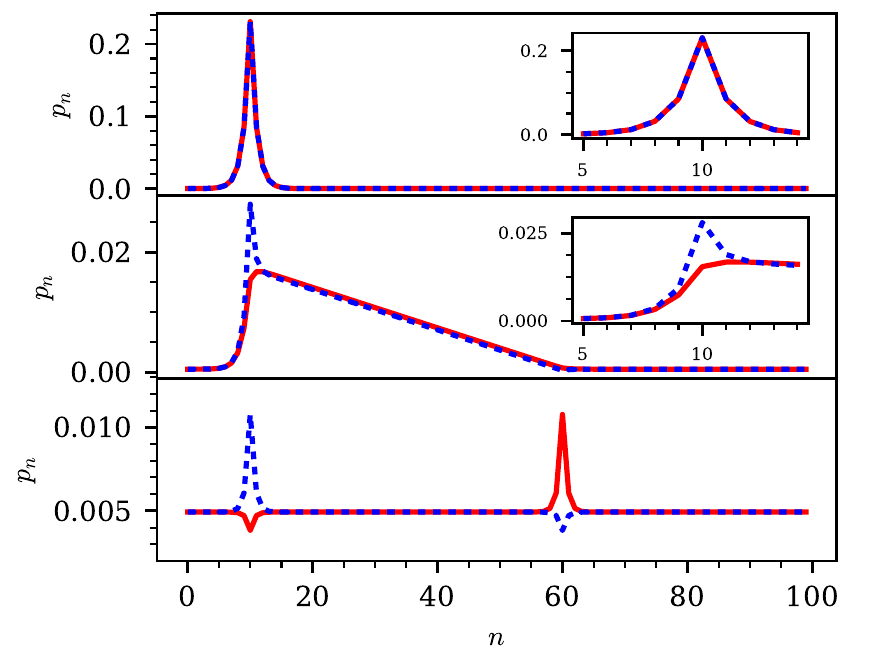} 
\put(19,69){(a)}
\put(19,48){(b)}
\put(19,27){(c)}
\end{overpic}

\caption{\label{f12} 
The NESS for a system that is composed of two regions.
The probabilities on the upper and lower chains 
are plotted in dashed-blue and solid-red lines.
{\bf (a)}~The two regions are with opposite~$f$. 
The sink interface is located at $n{=}10$,
while the other interface is at $n{=}60$.
{\bf (b)}~The ${[10,60]}$ region is with $\phi$ only, 
and the other region is with $f$ only.
{\bf (c)}~The two regions are with opposite~$\phi$.
Strong localization near the interface is observed only 
in case (a) where the topological index is non-zero. 
Zoom is provided in the panels of~(a) and~(b).
}
\end{figure}

If we have an interface between two regimes, that do not have the same ${ (f,\phi) }$ bias, 
there is still a possibility to observe interface-states, even if the topological-index is zero. 
This is demonstrated in \Fig{f12}. 
We see there that strong localization near the interface is observed 
only in panel (a) where the ${\delta w \ne 0}$. 
In panel (b) the high-probability interface and the low probability interface  
are mediated by a linear variation in the zero drift ($f=0$) region, 
as in Ohmic systems. In panel (c) the sink in one chain is in-fact a saddle, 
due to its coupling to the other chain, hence the localization is weak. 
If a longer sample is taken, the hump in panel~(c) is smeared out (not shown).

\section{Linear algebra of non-Hermitian matrices}
\label{sec:lalgebra}

A non-hermitian operator $\bm{A}$ has right-eigenvectors
that satisfy $\bm{A} \ket{x} = \lambda_x \ket{x}$.
Where $\ket{x}$ is chosen to have the normalization $\Braket{x}{x} = 1$.
These eigenvectors are in general non-orthogonal: $\Braket{x}{y} \ne 0$.
To any right-eigenvector we can associate left-eigenvector through the adjoint operator:
$\bm{A}^{\dag} \ket{\tilde{x}} = \lambda_x^{*} \!\ket{\tilde{x}}$.
The right- and left-eigenvectors form a bi-orthogonal set
and we choose the normalization of the left-eigenvectors such that $\Braket{\tilde{x}}{y} = \delta_{x,y}$. 
For a complete basis $\bm{1} = \sum_x \Ketbra{x}{\tilde{x}}$.
The matrix representation $B_{y,x}$ of an operator $\bm{B}$, 
in the basis $\ket{x}$ is defined via $\bm{B} \ket{y} = \sum_{x} B_{x,y} \ket{x}$. 
One deduces that $B_{x,y} = \BraKet{\tilde{x}}{\bm{B}}{y}$, 
and $\bm{B} = \sum_{x,y} \KetBra{x}{B_{x,y}}{\tilde{y}}$.

Given a non-hermitian matrix $\bm{H}_0$ and some perturbation $\bm{V}$,
we define the right and left unperturbed eigenvectors $\ket{n}$ and $\bra{\tilde{m}}$. 
In this basis $\bm{H}_0$ is diagonal with eigenvalues $\lambda_n^{(0)}$. 
The perturbed eigenvalues in second order are \cite{Sternheim1972}:
\begin{align}\label{eq:lambda-pert-second-order}
& \lambda_n = \lambda_n^{(0)} + \lambda_n^{(1)} + \lambda_n^{(2)}\\
& \lambda_n^{(1)} = \BraKet{\tilde{n}}{\bm{V}}{n}  \\
& \lambda_n^{(2)} = \sum_{m} \frac{ \BraKet{\tilde{n}}{\bm{V}}{m} \BraKet{\tilde{m}}{\bm{V}}{n} }{ \lambda_n^{(0)} - \lambda_{m}^{(0)} } \label{eq:second-term-pet}
\end{align}
Note that $V_{n,m}$ will take different forms depending on 
the normalization of the basis, while the product $V_{n,m}V_{m,n}$ 
is independent of normalization due to the bi-orthornormality.

\begin{figure}
\centering
\includegraphics[]{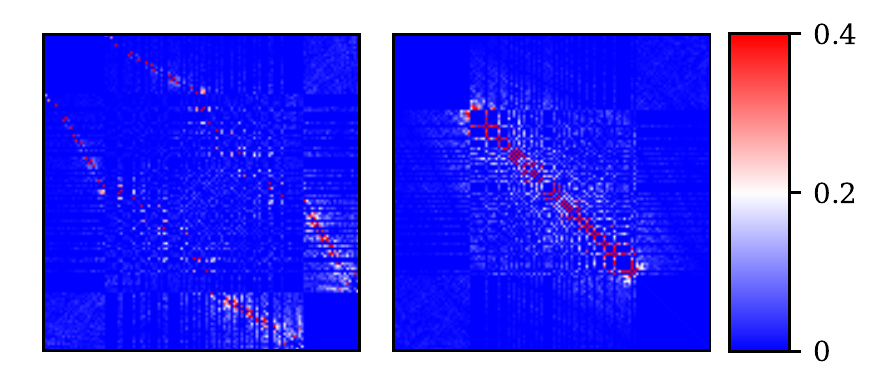} 
\caption{\label{fig:A-mat}
\textbf{Left panel}:
An image of $|A_{ij}A_{ji}^{*}|^{1/2}$ in the $\ket{\alpha,\pm}$ basis. 
The indices are sorted by $\lambda_{i}^{(0)}$. 
The propulsion is ${\bar{\phi}=2}$, and ${\sigma_{\phi} {=} 0.2}$, while ${\bar{f}=\sigma_f=0}$.
States that have similar $k$~content are strongly coupled.
If ${\bar{\phi}=0}$ (not displayed) all the couplings becomes 
comparable as implied by \Eq{eq:kAk}.  
\textbf{Right panel}: 
Here $\bar{\phi}=0$ and $\sigma_{\phi} {=} 1$. 
The elements are divided by the eigenvalue differences $|\lambda_{i}^{(0)}-\lambda_{j}^{(0)}|$.
The red color implies strong coupling that induce strong mixing. 
\textbf{Note:} In both figures diagonals are excluded. 
The color-scale in the left panel is arbitrary, while that of the right panel is absolute. 
}                              
\end{figure}

{The $\alpha$ eigenstates of \Fig{fd} are perturbed versions of $k$ eigenstates. 
On the basis of standard Fourier-analysis argumentation the uncertainty 
in~$k$ is determined by the inverse localization length~$1/L$. 
Given an $\alpha$ eigenstate, we can associate with it an average $k$~value. 
A demonstration for this insight is provided by the left panel of \Fig{fig:A-mat}.
It displays the magnitude of the anti-hermitian couplings between $\alpha$ states 
in the presence of propulsion with small-TSD. 
We see that those that have similar $k$-context are strongly coupled.  
If the average propulsion is switched off all the elements become comparable
in magnitude. Let us find an analytic estimate in the latter case.}

{The explicit expression for the anti-hermitian perturbation $\bm{A}$ 
is implied by inspection of $\bm{W}_{\text{hop}}$. Namely, 
\beq \nonumber 
\bm{A} = \sum_n \sinh\left(\frac{\phi_n}{2}\right)  
\Big[\Ketbra{n{+}1}{n} - \Ketbra{n}{n{+}1} \Big] 
\ \bm{\sigma}_z
\eeq
For the matrix elements of the anti-hermitian perturbation
we get ${\BraKet{k_2,-}{\bm{A}}{k_1,+}=A(q,\bar{k})}$, 
where ${q=k_1{-}k_2}$, and ${\bar{k}=(k_1{+}k_2)/2}$, and 
\beq \label{eq:kAk}
A(q,\bar{k}) = - \frac{i}{N} 
\sum_n 2 \sinh\left(\frac{\phi_n}{2}\right) e^{iq(n+(1/2))} \sin\left(\bar{k}\right)   
\ \ \ \ \ \ \ 
\eeq
We can use this expression in order to estimate 
the anti-hermitian couplings between $\alpha$ states
due to weak TSD.  Optionally we can calculate 
those  couplings numerically, as demonstrated 
in the right panel of \Fig{fig:A-mat}.  
Red color implies strong coupling that induces strong mixing.} 

{Eigenvalues in the $\re{[\lambda]} \in [2,4]$ window of an unbiased system 
become complex whenever the coupling $|A(q,\bar{k})|$ of \Eq{eq:kAk}
is bigger than (half) the eigenvalue difference between the states.
The sum in \Eq{eq:kAk} is the sum of $N$ independent random
terms, with zero average, hence for weak TSD, 
disregarding prefactor of order unity, 
\beq
\overline{|A(q,k)|} \ \approx \ \frac{\sigma_{\phi}}{\sqrt{N}} \left|\sin\left(k\right)\right|   
\eeq  
On the other hand, the spacings within $\re{[\lambda]} \in [2,4]$ is on the average $\propto 1/N$.
It follows that for large~$N$, the coupling to spacing ratio is $\propto \sqrt{N}$, 
leading to strong mixing and the emergence of complex eigenvalues, 
even if the TSD is very small.}


\newpage

\clearpage
\end{document}